\documentclass[ showpacs,aps]{revtex4-1} 
\usepackage{amsmath}
\usepackage{graphicx} 

\begin{document}

\title{\bf  Three-dimensional axisymmetric  sources  for  Majumdar-Papapetrou type  spacetimes }


\author{Gonzalo Garc\'{\i}a-Reyes}
\email[e-mail: ]{ggarcia@utp.edu.co}
\author{Kevin A.   Hern\'andez-G\'omez} 
\email[e-mail: ]{kevin\_201116@hotmail.com}

\affiliation{Departamento de F\'{\i}sica, Universidad Tecnol\'ogica de Pereira,
 A. A. 97, Pereira, Colombia}

\begin{abstract}

From Newtonian potential-density pairs we construct   three-dimensional
axisymmetric relativistic sources  for a Majumdar-Papapetrou type conformastatic
spacetime. As  simple examples,  we build   two family   of  relativistic thick disks  from of the  first two    Miyamoto-Nagai  potential-density pairs used
in Newtonian gravity to model flat galaxies, and a three-component relativistic model
of   galaxy (bulge, disk and dark matter halo). 
We study the equatorial
circular motion of test particles around such structures.   
Also the stability of the
orbits is analyzed for radial perturbation using an extension of the Rayleigh criterion.   In all examples, the relativistic effects
are analyzed and   compared with the Newtonian approximation.
The models considered satisfying all the energy conditions. 


\end{abstract}

\maketitle

\section{Introduction}

Axially  symmetric distribution of matter are important in astrophysics as models
of flat galaxies, accretion disks and certain stars, 
and in general relativity as sources of  vacuum gravitational fields.
In the context of  Newtonian gravity, a simple model of  highly flattened axisymmetric galaxies is  the Kuzmin's disk \cite{KUZMIN}.
This potential-density pair  is  first member of the Kuzmin-Toomre family of disks 
\cite{TOOMRE}.
These models are
constructed using the image method that is usually used to
solve problems in electrostatics.  Such  structures have no
boundary of the mass but as the surface mass density
decreases rapidly one can define a cutoff radius, of the
order of the galactic disk radius, and, in principle, to
consider these disks as finite. Three-dimensional models for flat galaxies were obtained by Miyamoto and Nagai  \cite{Miyamoto, Nagai} who thickened   all members of  Kuzmin-Toomre series of 
disk models.  The  models are a generalization of the
Plummer \cite{Plummer} and Kuzmin-Toomre models  and 
expresse the
mass components of a bulge and thin/thick disk of a galaxy.
 Its applications to models of 
Milky Way and other disk-like  galaxies are extensive  
\cite{Santillan,Nink1,Bajkova,Nink2}.
For other three-dimensional models of galaxies see for example \cite{Binney}. 

In curved spacetimes, exact solutions of Einstein field equation representing disk like configuration of matter also have been extensively studied.  In the case of thin disks, these solutions can be 
static or stationary and with or without radial pressure. 
Solutions for  static thin disks without radial pressure were first studied by  Bonnor and Sackfield \cite{BS},
and Morgan and Morgan \cite{MM1}, and with radial  pressure also by Morgan and Morgan
\cite{MM2}.   Several classes of exact solutions of the  Einstein field  equations
corresponding to static thin disks with or  without radial pressure have been
obtained by different authors
\cite{LP,CHGS,LO,LEM,BLK,BLP,GE}.
Rotating  thin disks that can be considered as a source of a Kerr metric were
presented by  Bi\u{c}\'ak and  Ledvinka \cite{BL}, while rotating disks with
heat flow were were studied by Gonz\'alez and Letelier \cite{GL2}.
The exact superposition of a disk and a static black hole was first considered
by Lemos and Letelier in Refs. \cite{LL1,LL2}. On the other hand,  relativistic thick disk models  were presented    in reference  \cite{G-L-thick}
and three-dimensional relativistic models of galaxies  for a Schwarzschild type conformastatic spacetime   in \cite{Let-galaxy}.   Also, relativistic  spherical sources  for a Majumdar-Papapetrou type conformastatic spacetime  were  studied  in \cite{Gon-shell}
and moreover   using  the   well-known ``displace, cut and reflect'' method were also  constructed there   models   
of thin disks surrounded of haloes of visible matter.

In this work, we construct    three-dimensional   axisymmetric sources  for a Majumdar-Papapetrou type conformastatic   spacetime from     more realistic Newtonian potential-density pairs (Poisson's  equation),
adding to the previous models  an  additional degree of reality to disks, its thickness, and the dark matter. In all examples,  the relativistic effects
are analyzed and  compared with the Newtonian approximation.
The models are illustrated  with  two     family  of  relativistic  thick disks  
built  from  the  first two   Miyamoto-Nagai  potential-density pair used  in the context of Newtonian galactic models, and a three-component relativistic model
of   galaxy (bulge, disk and dark matter halo).  In the latter case,
we use  as seed potentials the  Miyamoto-Nagai  potentials for the central bulge and the disk  and for the dark matter halo the well-known 
Navarro-Frenk-White (NFW) model \cite{NFW}.

The paper is organized as follows.  In Sect. II we present the method  to construct  different   three-dimensional axisymmetric configurations of matter from  a Newtonian potential-density pair in the particular case of  
 a  Majumdar-Papapetrou type   conformastatic spacetime. 
In Sects. III - V  the formalism is employed  to construct   two family of   relativistic thick disk and a three-component relativistic model of   galaxy.    In each case  the equatorial circular motion of test particles around the
structures is analyzed.  Also the stability of the orbits is studied  for radial perturbation using an extension of the Rayleigh criterion \cite{RAYL,FLU}.  Finally, in Sect. VI we summarize and discuss the results
obtained.



\section{Einstein equations  and motion of   particles} 
We consider a    conformastatic axisymmetric spacetime \cite{synge, kramer}
in cylindrical coordinates  ($t$, $\varphi$, $R$, $z$) and in the particular form
\begin{equation}
ds^2 =  - \left( 1- \frac{\phi}{2} \right)^{-4} dt^2 + \left( 1- \frac{\phi}{2} \right)^{4} (R^2 d \varphi^2  + dR^2 + dz^2),  \label{eq:met}
\end{equation}
where $\phi$ is function of $R$ and $z$ only.   Due to its known use  in context of electrostatic fields  \cite{majum, papa}  these fields   can be called  Majumdar-Papapetrou type  espacetimes.

   The Einstein field equations $G_{ab}= 8 \pi G  T_{ab} $ yield the following non-zero  components of the energy-momentum tensor  
\begin{subequations}\begin{eqnarray}
T^t \ _t  & = & - \frac{\nabla ^2 \phi }{4 \pi G   \left( 1- \frac{\phi}{2} \right)^{5} },
   \label{eq:Ttt}  \\
&  &  \nonumber \\
T^\varphi \ _\varphi &=&   \frac{\phi _{,R}^2 + \phi _{,z}^2 }{8 \pi G \left( 1- \frac{\phi}{2} \right)^{6}},   \label{eq:T11} \\
& &  \nonumber  \\
T^z \ _z  & = &  - T^R \ _R =   \frac{\phi _{,R}^2 - \phi _{,z}^2 }{8 \pi G \left( 1- \frac{\phi}{2} \right)^{6}}  .   \label{eq:TRR} \\
&  &  \nonumber \\
T^R \ _z  & = &  -  \frac{\phi _{,R}  \phi _{,z}}{ 4 \pi G \left( 1- \frac{\phi}{2} \right)^{6}}.   \label{eq:TRz}
\end{eqnarray} \end{subequations} 

Now, in order to analyze the matter content of the disks is   necessary to
compute the eigenvalues and eigenvectors of  the energy-momentum tensor. The
eigenvalue problem for the   energy-momentum tensor   (\ref{eq:Ttt}) - (\ref{eq:TRz})
has the solutions
\begin{subequations}\begin{eqnarray}
\lambda_0 \  &=& \  T^t \ _t  ,  \\
\lambda_1 \  &=& \  T^\varphi \ _\varphi,  \\
\lambda_2 \  &=&  - \lambda_3  = \sqrt{ ( T^R \ _R)^2  + ( T^R \ _z)^2  }.
\end{eqnarray} \end{subequations} 
The corresponding eigenvectors are
\begin{subequations}\begin{eqnarray}
{\bf V} & = & ( 1- \frac{\phi}{2} )^{2} ( 1, 0, 0, 0 ),	\\
	&	&  \nonumber	\\
{\bf X} & = & \frac{( 1- \frac{\phi}{2} )^{-2}}{ R  } ( 0, 1, 0, 0 ),  \\
	&	&  \nonumber	\\
{\bf Y} & = &  \frac { ( 1- \frac{\phi}{2} )^{-2} T^R \ _z }   
{    \left [ 2 D (  - T^R \ _R + D ) \right ]^{1/2}  }  (0  , 0 , 1,  \frac { -  T^R \ _R  + D }{ T^R \ _z}   ),  \\
	&	&  \nonumber	\\
{\bf Z} & = &  \frac { ( 1- \frac{\phi}{2} )^{-2} T^R \ _z }   
{    \left [ 2 D (  T^R \ _R + D ) \right ]^{1/2}  } (0  , 0 , 1,  \frac { -  T^R \ _R  - D }{T^R \ _z}  ),
\end{eqnarray} \end{subequations} 
where
\begin{equation}
D = \sqrt{ ( T^R \ _R)^2  + ( T^R \ _z)^2} .
\end{equation}

In terms of the orthonormal  tetrad or comoving observer  ${{\rm e}_{ (a)}}^b = \{
V^b , X^b,Y^b,  Z^b \}$,  the energy density is given by 
\begin{equation}
\rho= - \lambda_0 =  \frac{\nabla ^2 \phi }{4 \pi G   \left( 1- \frac{\phi}{2} \right)^{5} } ,   \label{eq:rho}
\end{equation}
and the stresses (pressure or tensions)  by  $p_i = \lambda _i $.

The expression   for the relativistic energy density   (\ref{eq:rho}) is a Poisson type nonlinear equation which can be solved by guessing the function $\phi$. The stresses follow directly.
In the Newtonian limit  $\phi \ll 1$,   the
relativistic energy density  $\rho$  must reduce to Poisson's
equation
\begin{equation}
 \nabla ^2 \Phi = 4 \pi G \rho_N.  \label{eq: poisson}
\end{equation}
This condition is satisfied by taking  $ \phi =  \Phi$.

Thus,  the physical quantities associated with the matter distribution are given by  
\begin{subequations}\begin{eqnarray}
\rho& = & \frac{ \rho_N } {  (1- \frac{\Phi}{2} )^5 }    ,     \label{eq: en} \\ 
&  & \nonumber   \\
 p_\varphi   & = &  p_R = -p_z   =  \frac{\nabla \Phi \cdot  \nabla \Phi}{8 \pi G \left( 1- \frac{\Phi}{2} \right)^{6}},  \\
&  &  \nonumber 
\end{eqnarray} \end{subequations} 
and  the average pressure  by  
\begin{equation} 
p= \frac 1 3 (p_\varphi  + p_R + p_z ) =  \frac{\nabla \Phi \cdot  \nabla \Phi}{24 \pi G \left( 1- \frac{\Phi}{2} \right)^{6}} .
\end{equation}

Moreover, in order to have a physically meaningful matter distribution the components of the energy-momentum tensor must  satisfy the energy conditions. The weak energy condition
requires that   $\rho\geq 0$, whereas  the dominant energy condition states that 
$|\rho|  \geq |p_i|$. The strong energy
condition  imposes
the condition  
$\rho_{eff} = \rho+ p_r + p_\theta + p_\varphi \geq 0$,  where $\rho_{eff}$
is the ``effective Newtonian density''. 



A useful  parameter related  to the motion of test  particles around the structures on the
equatorial plane  is circular speed $v_c$  (rotation curves).
For circular, equatorial  orbits the 4-velocity  ${\bf u}$  of the particles with respect to the coordinates
frame has  components ${\bf u} = u^t(1,\omega , 0,0)$, where  $\omega= u^\varphi/u^t=\frac{d
\varphi}{d t}$ is  the  angular speed of the test particles. 
With respect to tetrad the 4-velocity  has component
\begin{equation}
u^{(a)} = e^{(a)}_{\ \ \ b} u^b,
\end{equation}
and  the  3-velocity 
\begin{equation}
v^{(i)} = \frac { u^{(i)} } { u^{(t)} }  =   \frac { e^{(i)}_{\ \  \ a} u^a }{  e^{(t)}_
{ \ \ \ a}  u^a  }. 
\end{equation}

For circular, equatorial  orbits   the
only nonvanishing velocity  component is $v^{ (\varphi)}$, and is given by
\begin{equation}
 [v^{ (\varphi)}]^2 = v_c^2= - \frac{ g_{\varphi \varphi} }{ g_{tt} } \omega ^2 ,  \label{eq:vc2}
\end{equation}
and  represents   the circular speed   (rotation profile)   of the particle as seen by an observer at infinity.

On the other hand, the angular speed  $\omega$ can be calculated  considering  the motion of the particles along geodesics.  For the spacetime  (\ref{eq:met}),
the radial motion's equation  is given by 
\begin{equation}
 g_{ab, R}u^a u^b = 0,
\label{eq:geo}
\end{equation}
to obtain 
\begin{equation}
\omega ^2 = - \frac{g_{t t, R}}{g_{\varphi \varphi, R }}.
\end{equation}
In addition,  $u^t$ obtains normalizing  $u^a$, that is requiring  $g_{ab}u^au^b=-1$, so that
\begin{equation}
 (u^t)^2 = - \frac {1}{g_{\varphi \varphi} \omega^2 + g_{tt}} \label{eq:u0}.
\end{equation}

Thus,  the circular speed is given by
\begin{equation}
v_c^2 = \frac { v_{N}^2}  {1- \frac{\Phi}{2} - v_{N}^2   } ,  \label{eq:vc} 
\end{equation}
where
\begin{equation} 
v_{N}^2 = R \Phi_{,R}   \label{eq:vn2} 
\end{equation}
is the Newtonian circular speed.

To  analyze the stability of the particles against radial
perturbations we can use an extension of the Rayleigh criteria of stability of a fluid in rest in a
gravitational field \cite{FLU}
\begin{equation}
\frac{d(h^2)}{dR} \ > \ 0 ,
\end{equation}
where $h$  is the specific
angular momentum,   defined as $h =
g_{\varphi a} u^a$. For  circular, planar orbits  we obtain 
\begin{equation}
h^2 = \frac{R^2 \left( 1- \frac{\Phi}{2} \right)^{4}  v_c ^2}{1- v_c ^2} .
\label{eq:h}
\end{equation}
All above quantities are evaluated on the equatorial plane $z = 0$.


\section{ First family of Miyamoto-Nagai-like  thick disks } 

Models of thick disks in Newtonan gravity  can be obtained by a 
Miyamoto-Nagai transformation used to
generate three-dimensional potential-density
pairs from the Kuzmin-Toomre
thin disks.  This procedure is equivalent to change in the   Kuzmin-Toomre disks
$ |z| \rightarrow \sqrt{z^2 + b^2} $, where
$b$ is a positive parameter.
The models describe the stratification of mass in the central bulge as well in the disk part of galaxies. Thus, when this transformation is applied to the   zeroth order Kuzmin-Toomre disk we obtain the  first   Miyamoto-Nagai  potential-density pair
\begin{subequations}\begin{eqnarray}
\Phi &=& -\dfrac{GM}{\sqrt{R^2 + \left(a +\sqrt{ z^2 + b^2}\right)^2}} ,
\label{eq:PhiMN}  \\
&  & \nonumber   \\
\rho_N &= & \frac{ b^2 M
\left[   a R^2 + ( a + 3 \sqrt{z^2 + b^2 }  ) ( a +  \sqrt{z^2 + b^2 } )^2 \right]   }
{ 4\pi \left[ R^2+\left( a+\sqrt{z^2 + b^2 }\right) ^2 \right]^{5/2}    ( z^2 + b^2 )^{3/2} } , \label{eq:rhoMN}
\end{eqnarray} \end{subequations} 
where the parameter $a$,  $b$  and $M$ are the length, height scales and the  mass of the  disk like distribution.    The Newtonian circular speed on the equatorial plane  is given by
\begin{equation}
v_{N}^2 = \frac{GM R^2}{\left[  R^2 +  (a + b )^2  \right]^{3/2} }.  
\label{eq:vMN}
\end{equation}

The relativistic expressions for the energy density and average pressure are 
\begin{subequations}\begin{eqnarray}
\rho &=&  \frac{  b^2 
\left[  a \tilde R^2 + ( \tilde a + 3 \tilde \zeta ) (\tilde a +  \tilde \zeta)^2 \right]   }
{ 4 \pi  G^3 M^2 \left[  \sqrt {  \tilde R^2+\left( \tilde a+\tilde \zeta\right) ^2 } + 1/2 \right ]^ 5    \tilde \zeta^3  } ,  \\
p  &= & \dfrac{\left(\tilde{a}+ \tilde \zeta\right)^2  \tilde{z}^2 + \tilde \zeta ^2 \tilde{R}^2}{24\pi G^3 M^2 \left[  \sqrt { \tilde{R}^2 + (\tilde{a}+ \tilde \zeta)^2  }
 +  1/2 \right] ^6 \tilde \zeta^2 } ,
\end{eqnarray} \end{subequations} 
where $ \tilde{R} = R/GM $, $ \tilde{z} = z/GM $, $ \tilde{a} = a/GM $, $ \tilde{b} = b/GM $, and $ \tilde \zeta = \sqrt{\tilde{z}^2 + \tilde{b}^2 }$. We note that the energy density is always  
a positive quantity  in agreement with  the  weak energy condition and  
the stresses are positive  for all  values of parameters, so that we have always pressure.   In turn,
the circular speed and the specific angular momentum on $z=0$ are given by 
\begin{subequations}\begin{eqnarray}
v_c^2 & = & \dfrac{2 \tilde R^2}{2  \left ( \tilde R^2 + \tilde d ^2 \right )^{3/2}  - \tilde R^2
+ \tilde d ^2 },  \\
&  & \nonumber   \\
h ^2 & = & \frac { G^2 M^2 {\tilde R}^{4} \left[ 2 \sqrt { \tilde R ^2 + \tilde d ^2  }+1 \right] ^{4}}{ 8 \left( {\tilde R}^{2}+ \tilde d^2  \right) ^{2} 
\left[  2 \left ( \tilde R ^2 + \tilde d ^2   \right ) ^{3/2} - 3 \tilde R ^2 +  \tilde d^2   \right ] },
\end{eqnarray} \end{subequations} 
where $\tilde d = \tilde a + \tilde b$. 

\begin{figure}
$$
\begin{array}{cc}
\includegraphics[width=0.33\textwidth]{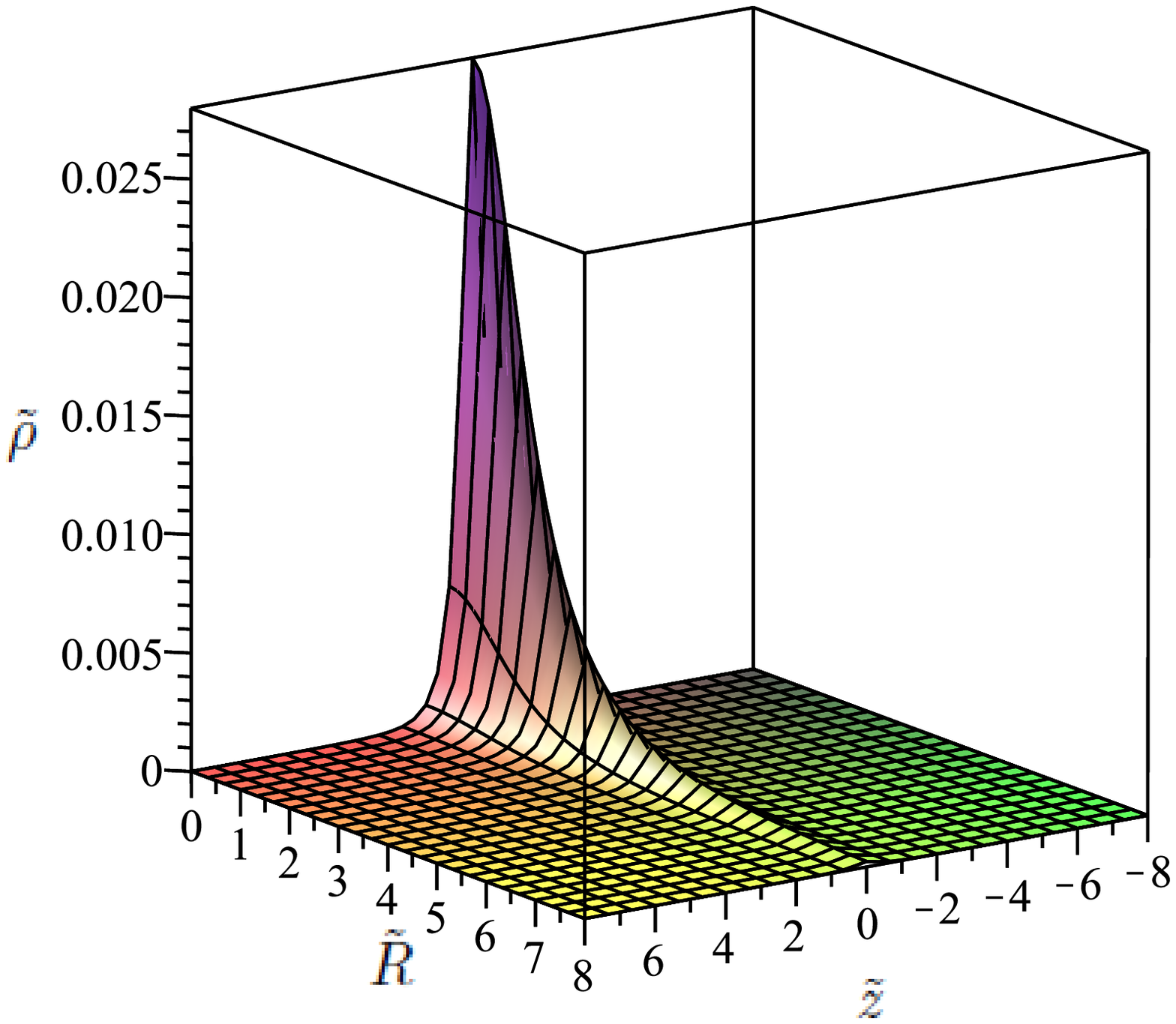} &   
\includegraphics[width=0.25\textwidth]{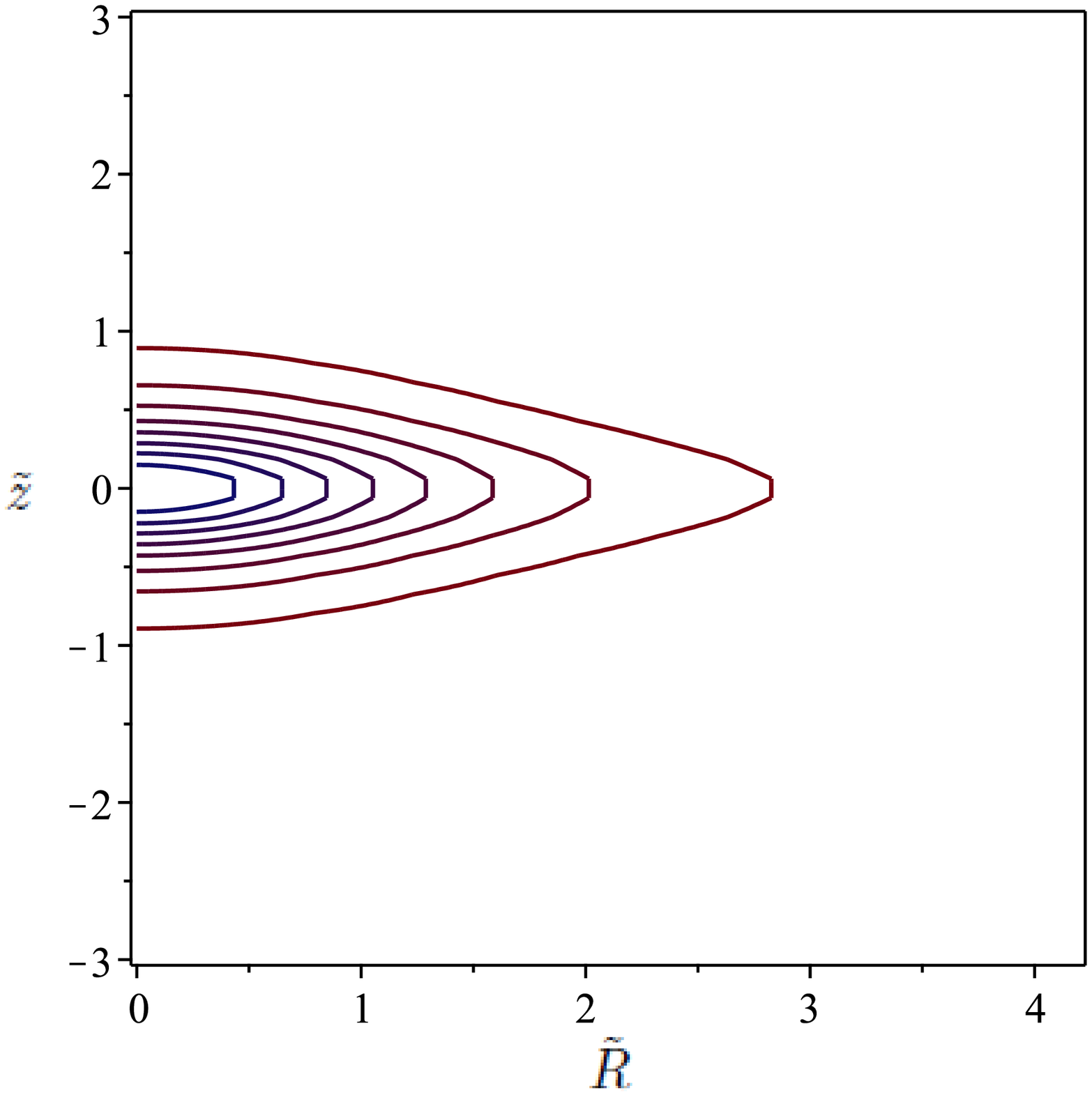}  \\  
 &  \\
(a)    &  (b)  \\
\includegraphics[width=0.33\textwidth]{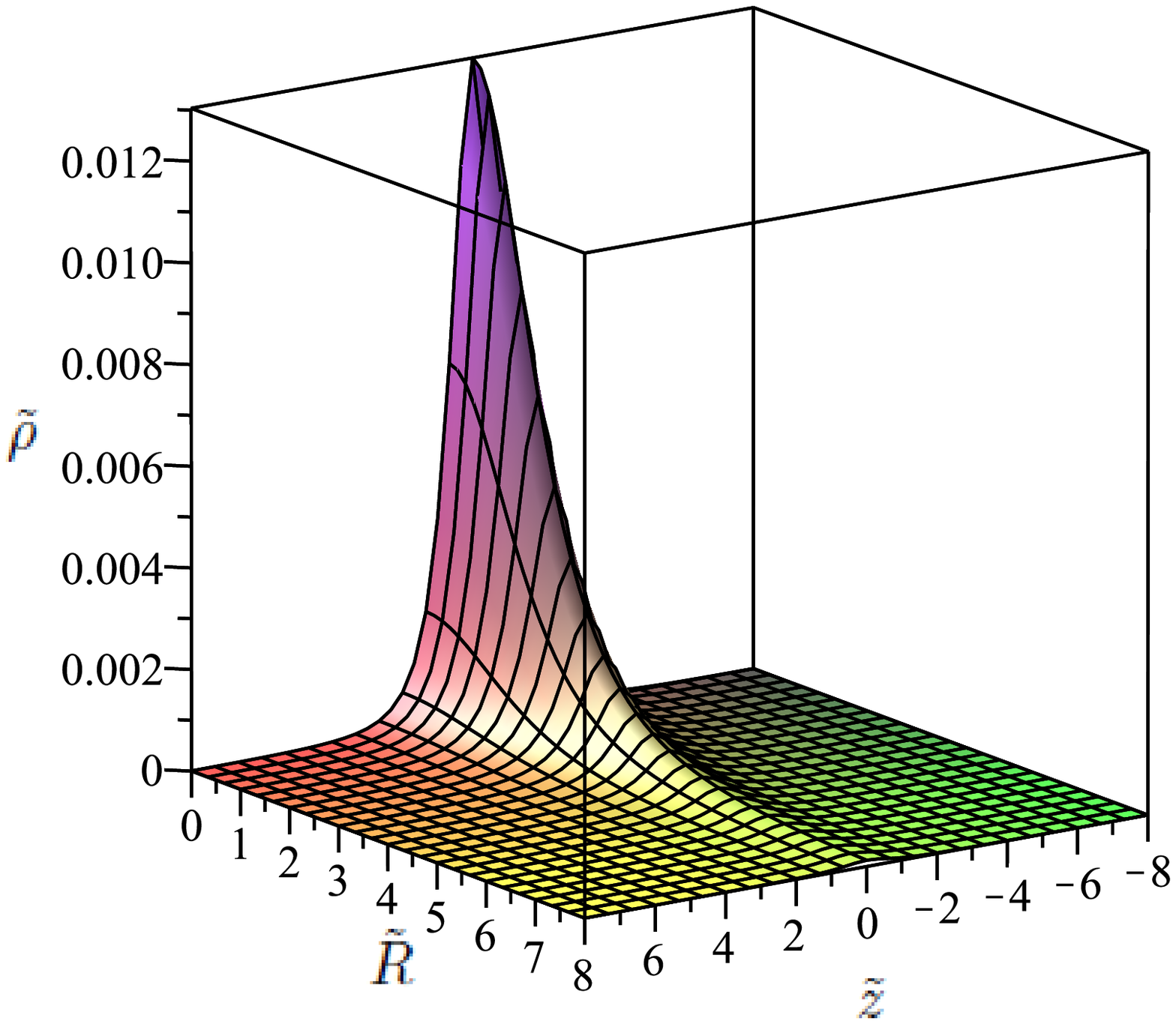} &   
\includegraphics[width=0.25\textwidth]{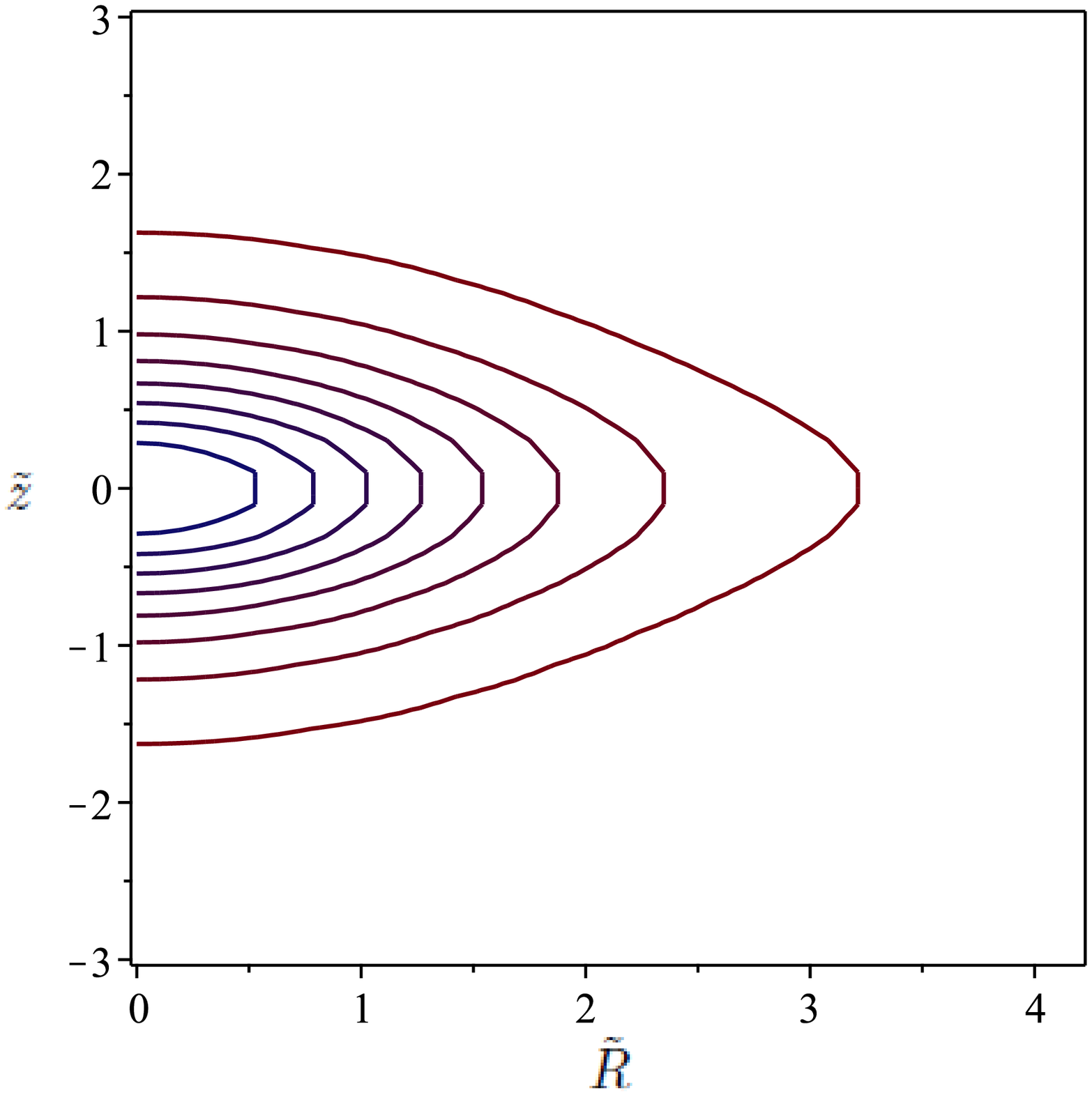}  \\  
 &  \\
(c)    &  (d)  \\
\includegraphics[width=0.33\textwidth]{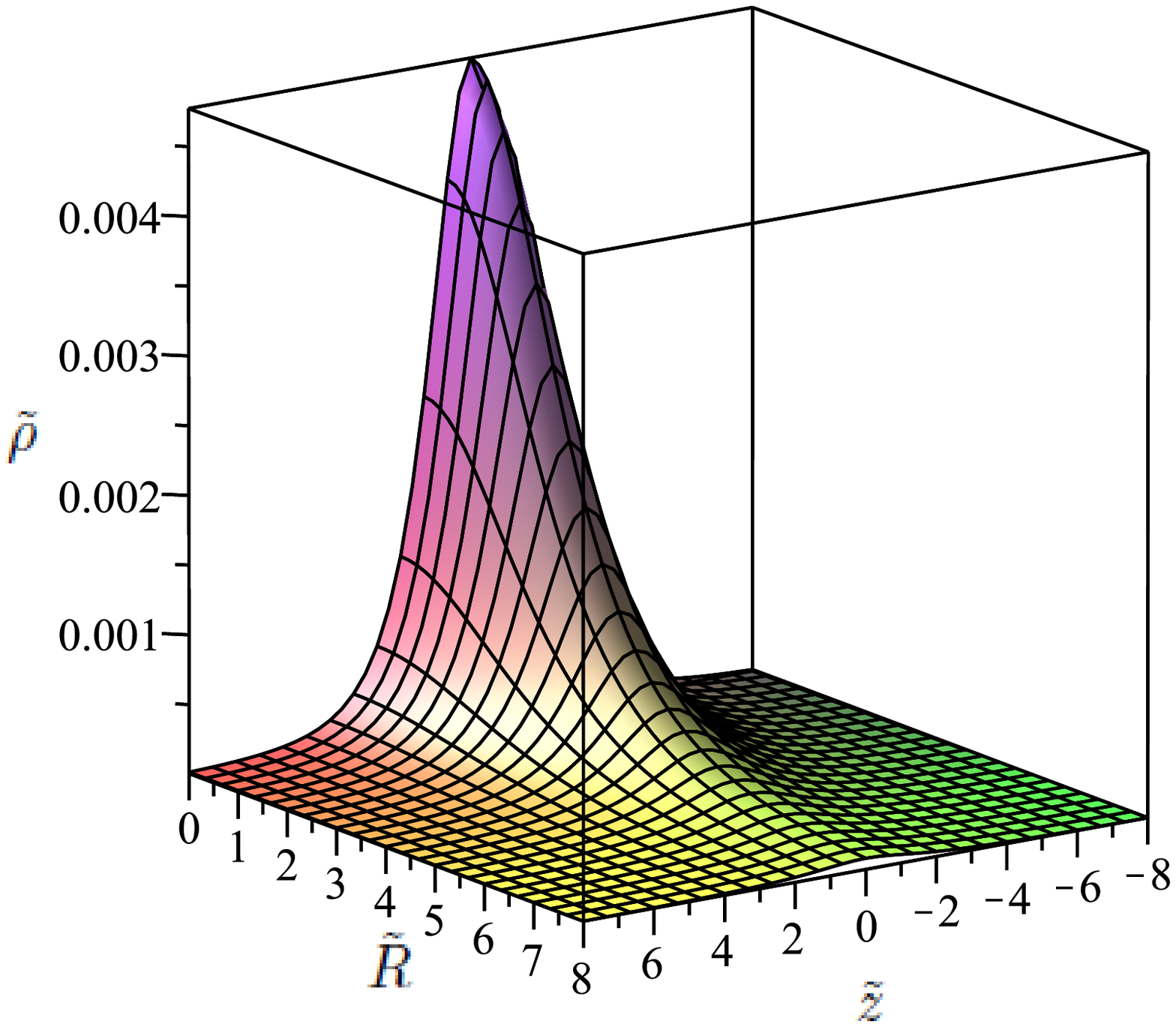} &   
\includegraphics[width=0.25\textwidth]{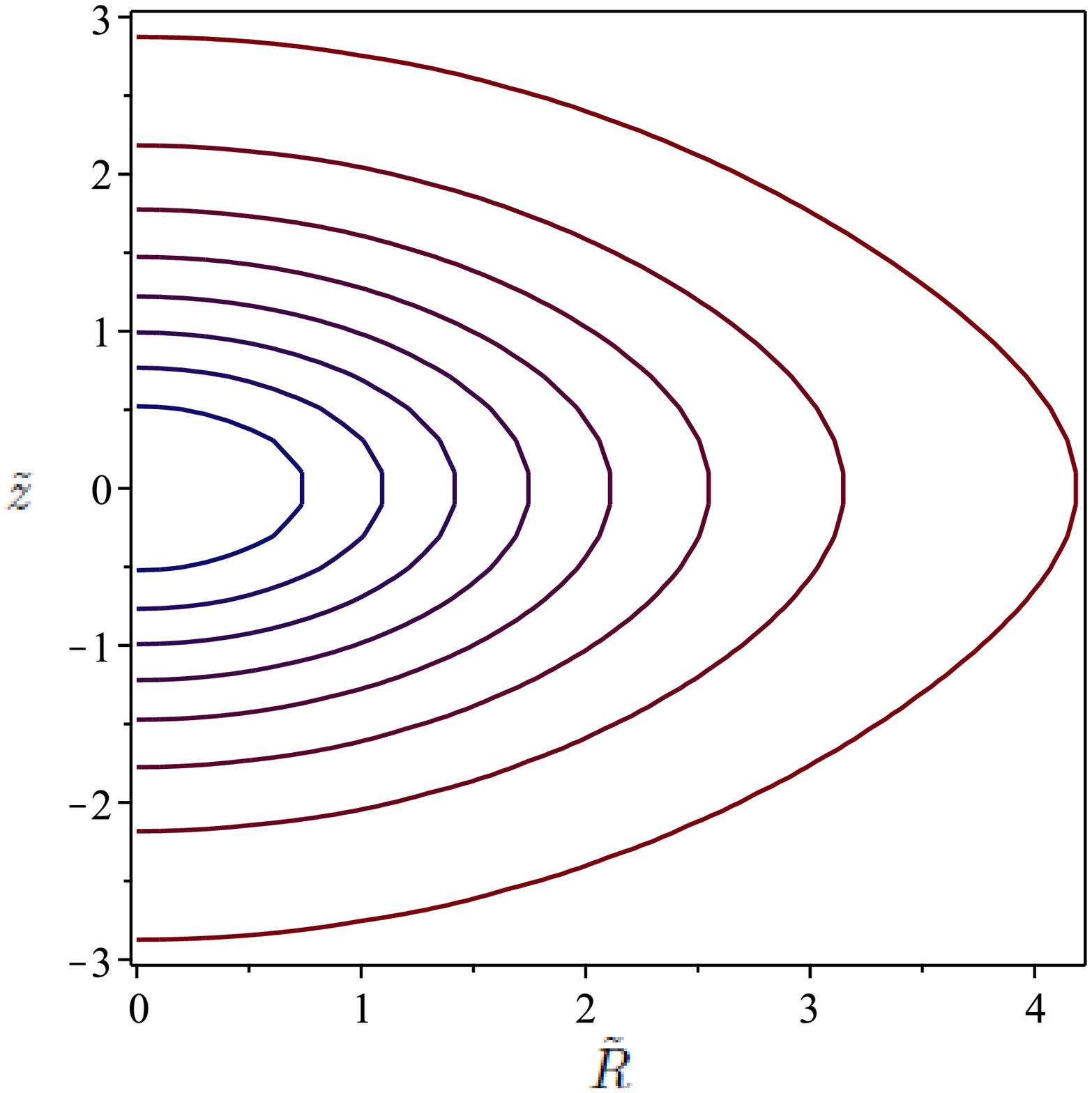}  \\  
&  \\
(e)    &  (f)
\end{array}
$$	
\caption{ The relativistic energy density $\tilde \rho$ and the isodensity  curves for the first model of Miyamoto-Nagai-like  thick disks  with parameters
 $\tilde a = 1$ and $\tilde b = 0.5$ (top figures),   $1$ (middle figures),  $ 2$ (bottom figures).} 
\label{fig:fig1}
\end{figure}

In figure  \ref{fig:fig1}  we plot,  as functions of $\tilde R$ and $\tilde z$,  the relativistic  energy density  $\tilde \rho=  G^3 M^2 \rho$   and the isodensity  curves  for the  first model
of  Miyamoto-Nagai-like thick disks
with parameters
 $\tilde a = 1$ and $\tilde b = 0.5$,   $1$, $ 2$.
 We see that the energy density  presents a   maximum  at the center of the distributions of matter, 
 and  then decreases rapidly with $\tilde R$    which  permits
to define a cut off radius $\tilde R_c$ and, in principle, to considerer the structures as  compact objects. Furthermore, as in the  Newtonian case,  as the ratio $\tilde b / \tilde a$
decrease the distribution of energy  becomes flatter so that $\tilde b / \tilde a $ is also  a 
measure of flatness of the models. 

\begin{figure}
$$
\begin{array}{cc}
\includegraphics[width=0.33\textwidth]{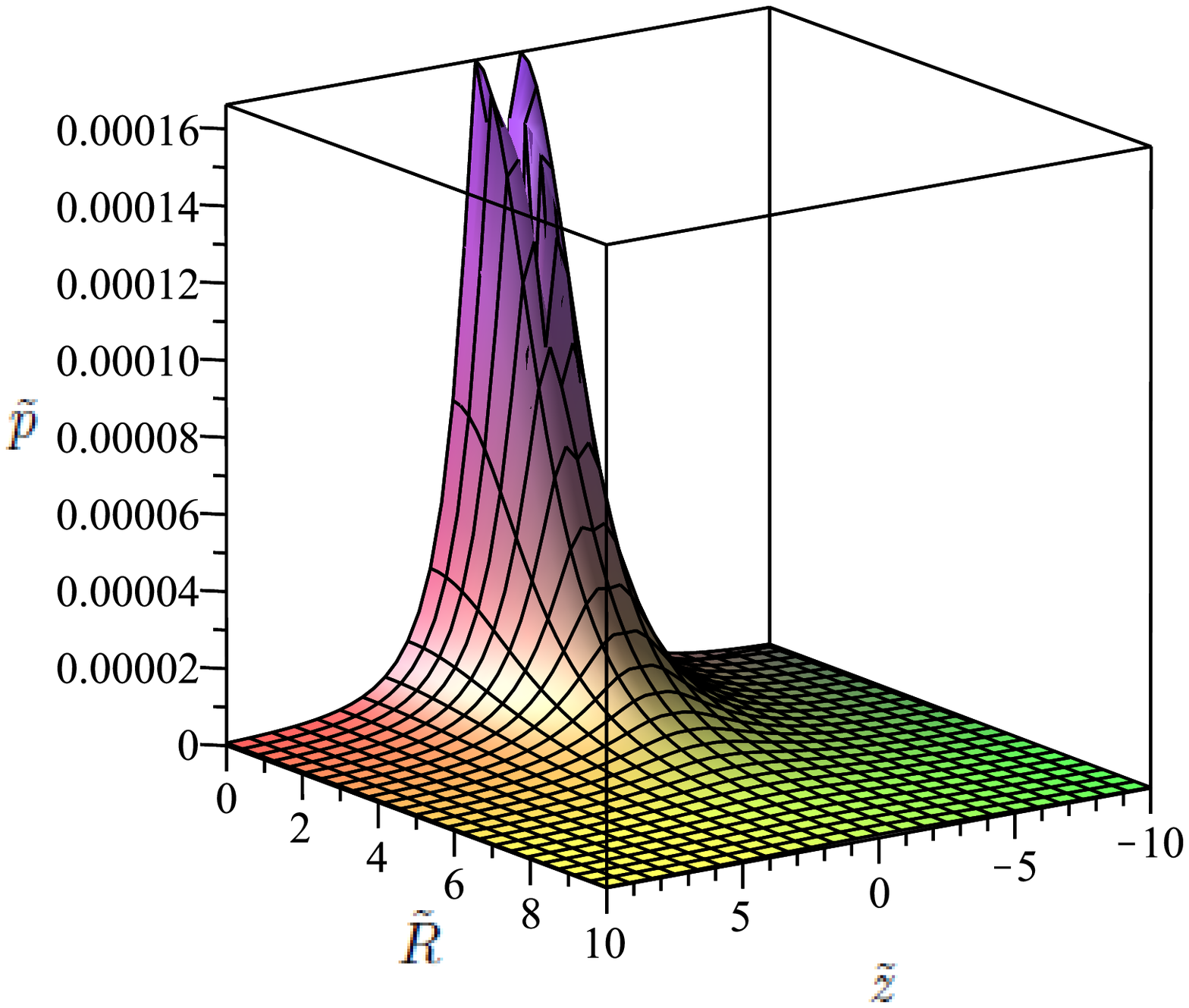} &   
\includegraphics[width=0.25\textwidth]{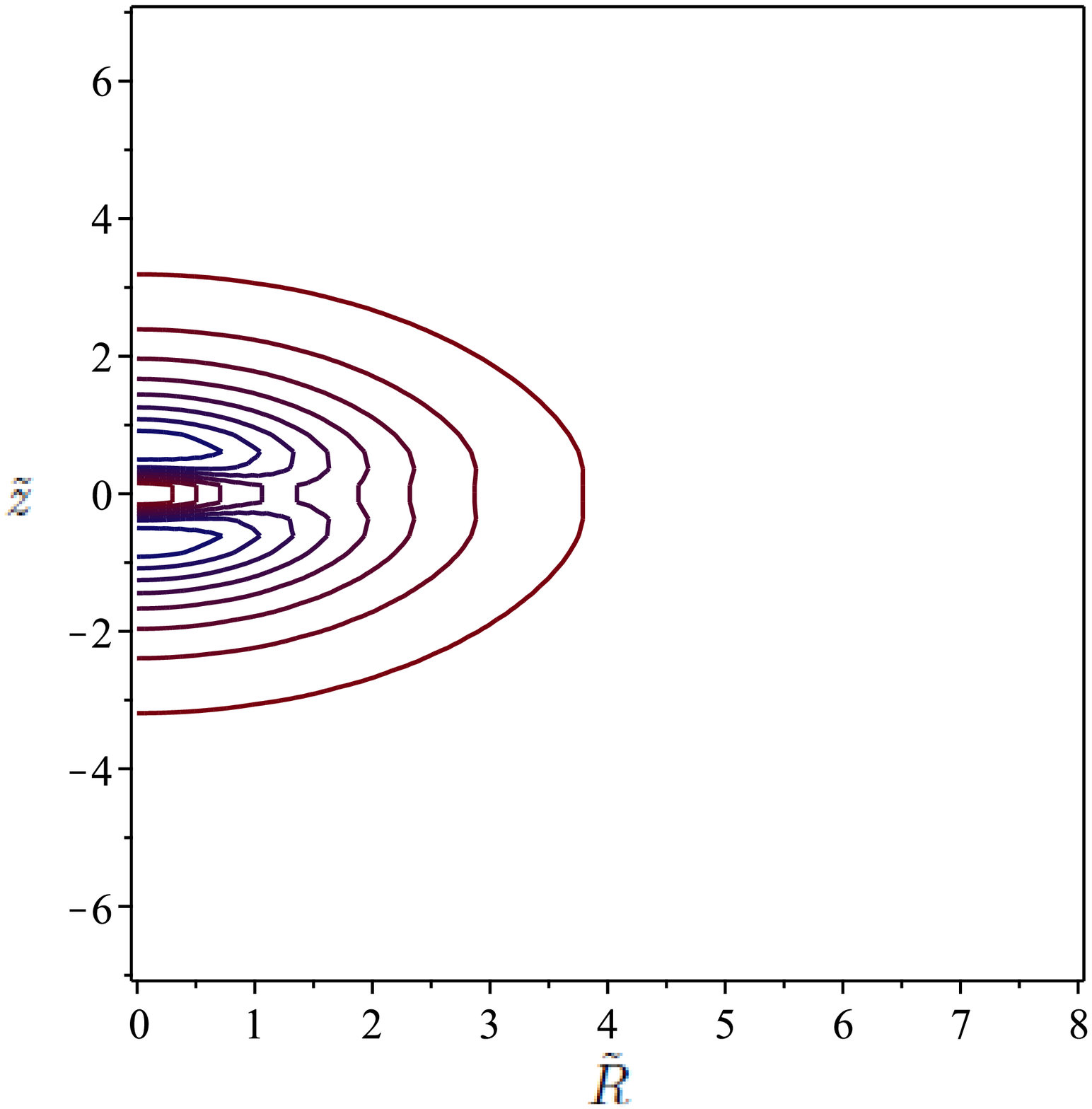}  \\  
 &  \\
(a)    &  (b)  \\
\includegraphics[width=0.33\textwidth]{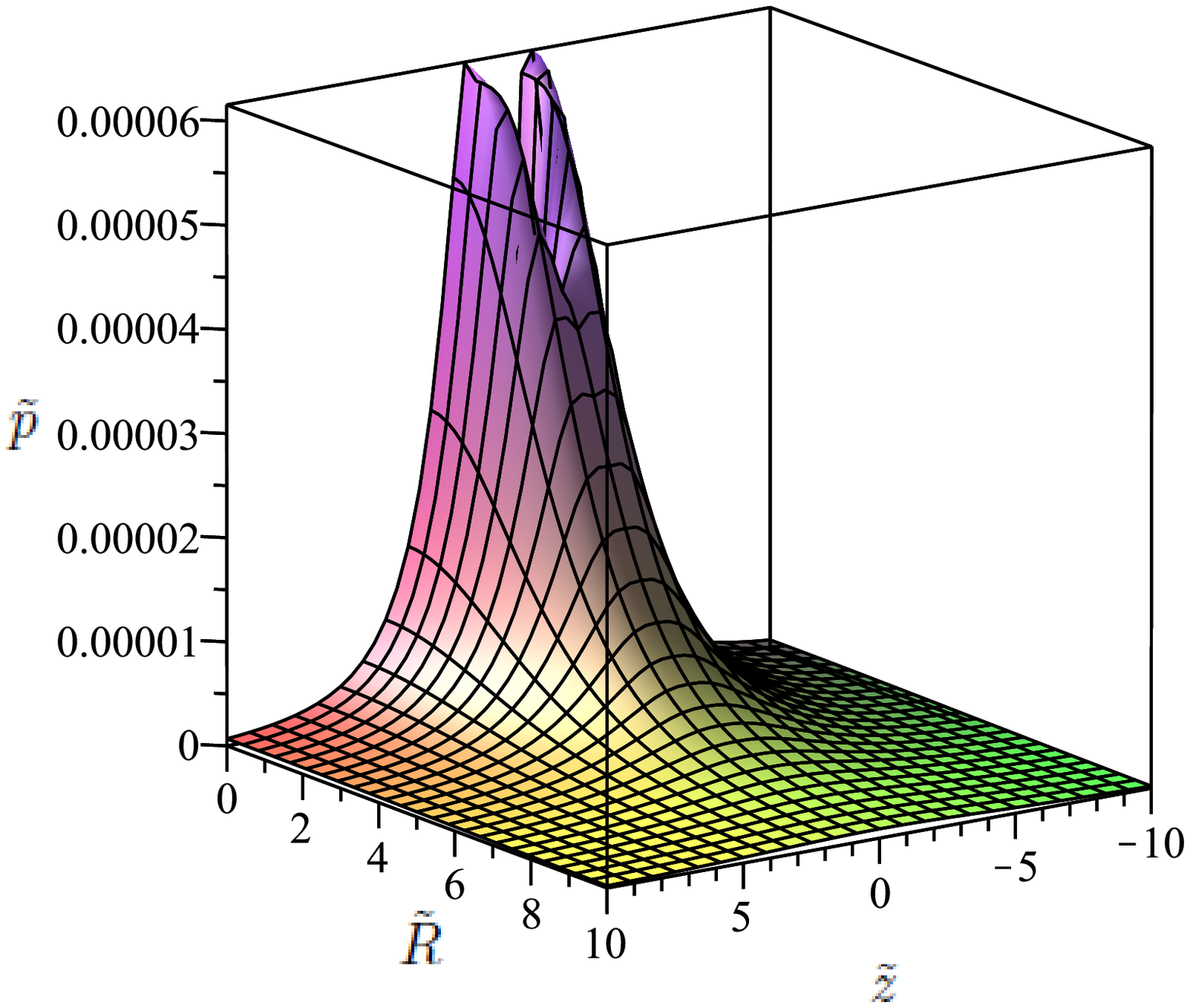} &   
\includegraphics[width=0.25\textwidth]{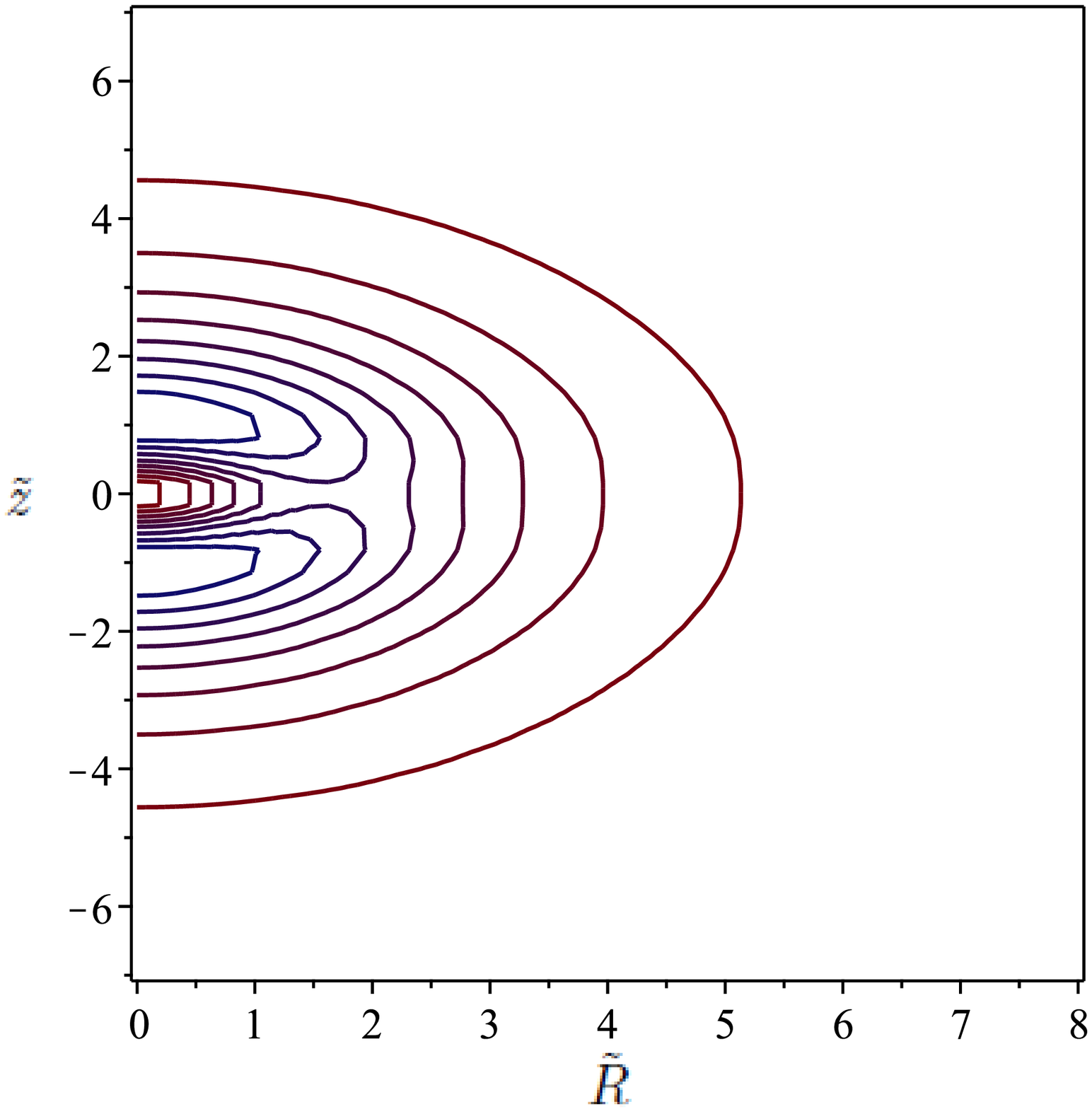}  \\  
 &  \\
(c)    &  (d)  \\
\includegraphics[width=0.33\textwidth]{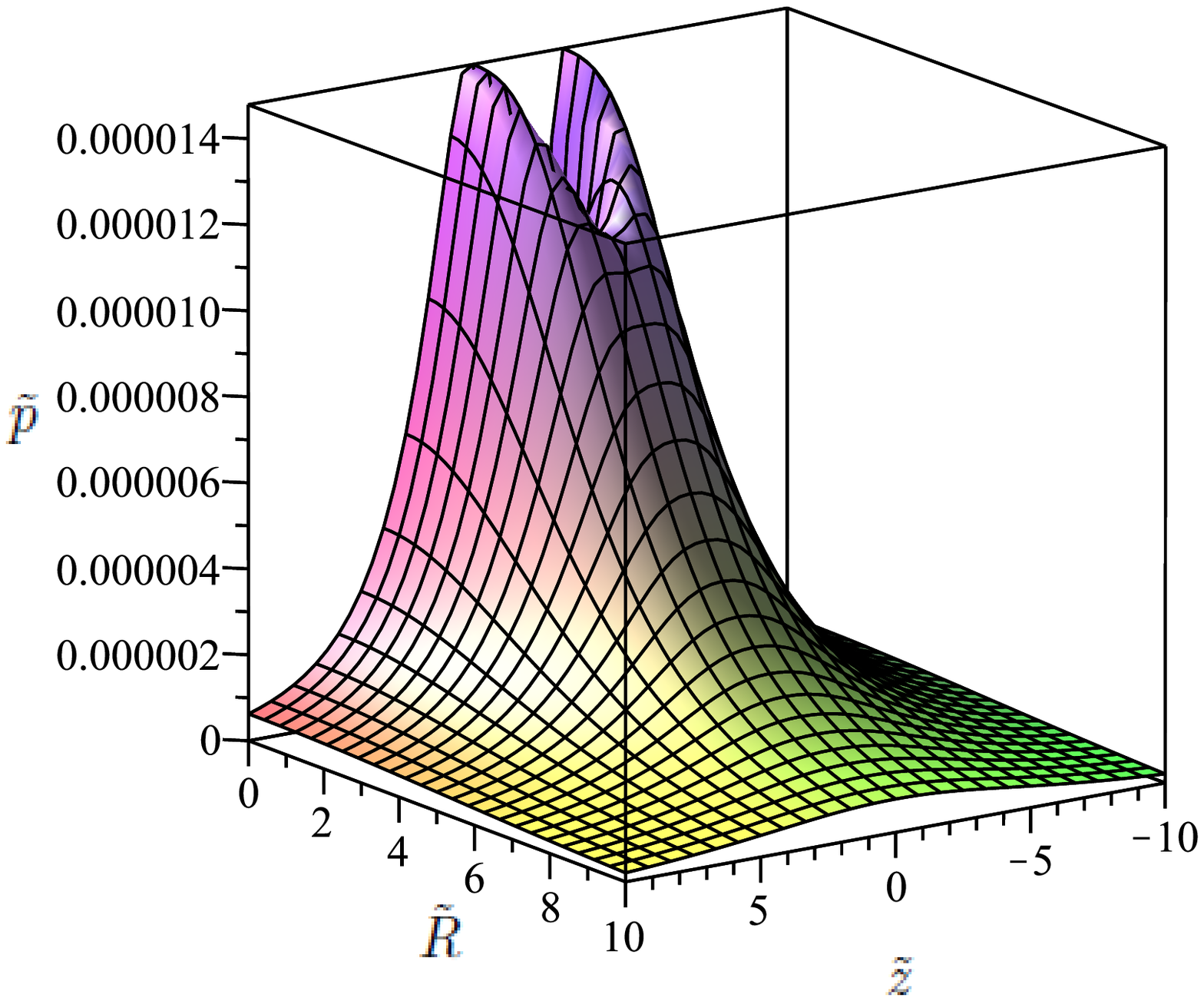} &   
\includegraphics[width=0.25\textwidth]{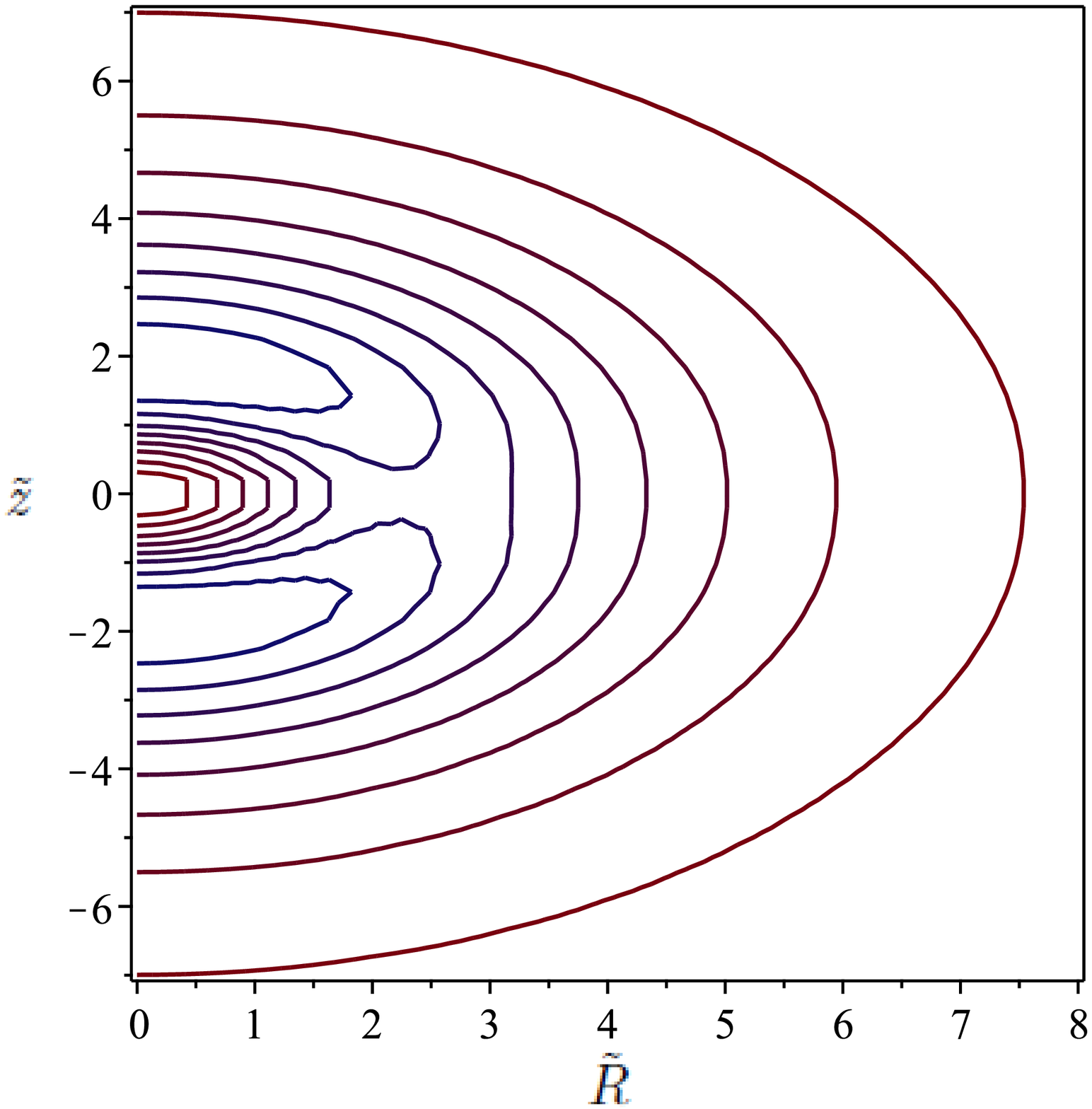}  \\  
&  \\
(e)    &  (f)
\end{array}
$$	
\caption{ The relativistic  average pressure $\tilde p$ and the level  curves  for the first model of Miyamoto-Nagai-like  thick disks  with parameters
 $\tilde a = 1$ and $\tilde b = 0.5$ (top figures),   $1$ (middle figures),  $ 2$ (bottom figures).} 
\label{fig:fig2}
\end{figure}

In figure \ref{fig:fig2}  we have  plotted   the average  pressure $\tilde p=  G^3 M^2  p$  and its level curves  for the same value of parameters. We observer that it vanishes at the center of the matter distributions, reach a region where its value is  maximum,
and then,  just like the  energy density,  vanishes fast with the increase of $\tilde R$.

\begin{figure}
$$
\begin{array}{cc}
\includegraphics[width=0.3\textwidth]{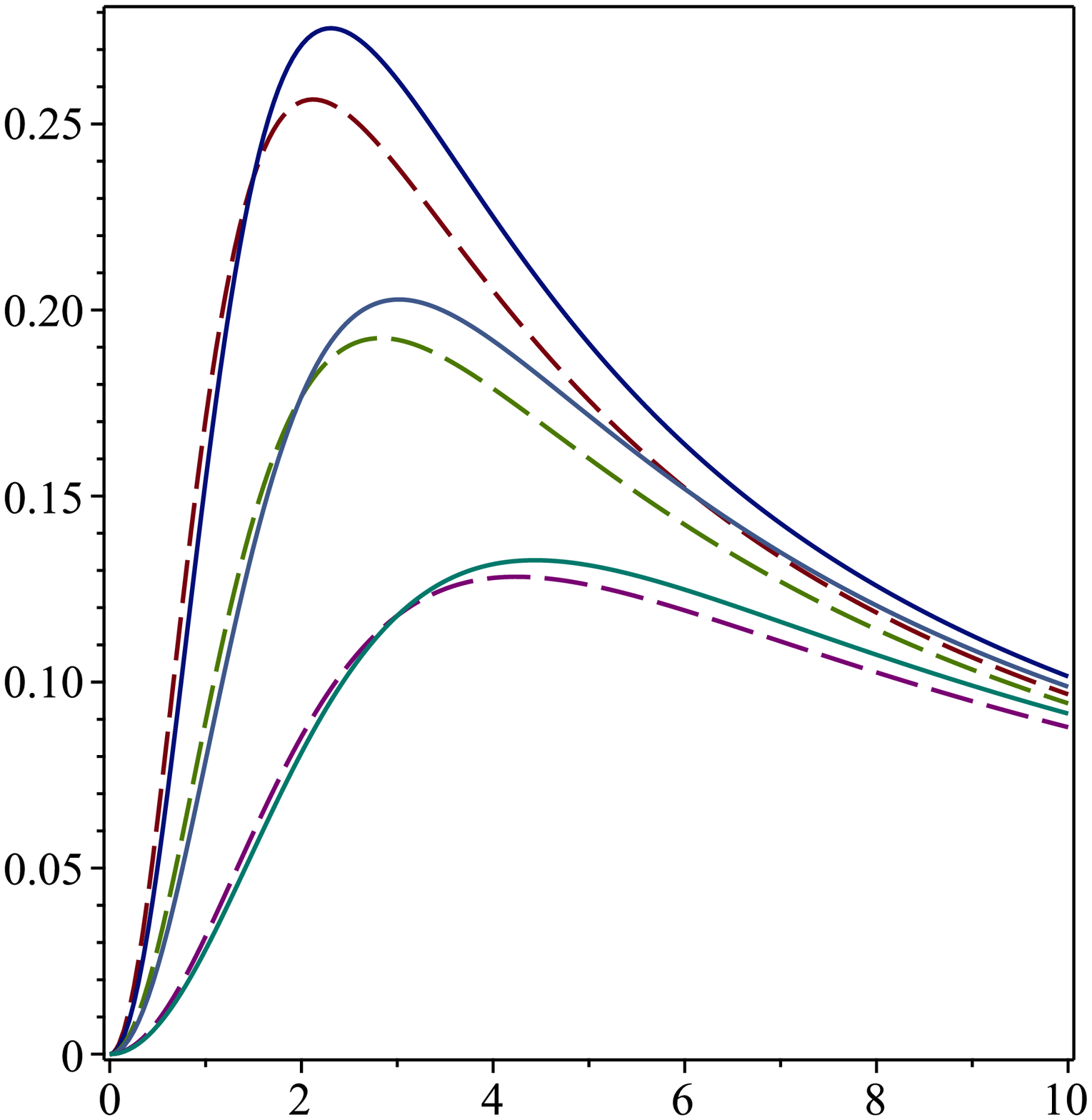} &   
\includegraphics[width=0.3\textwidth]{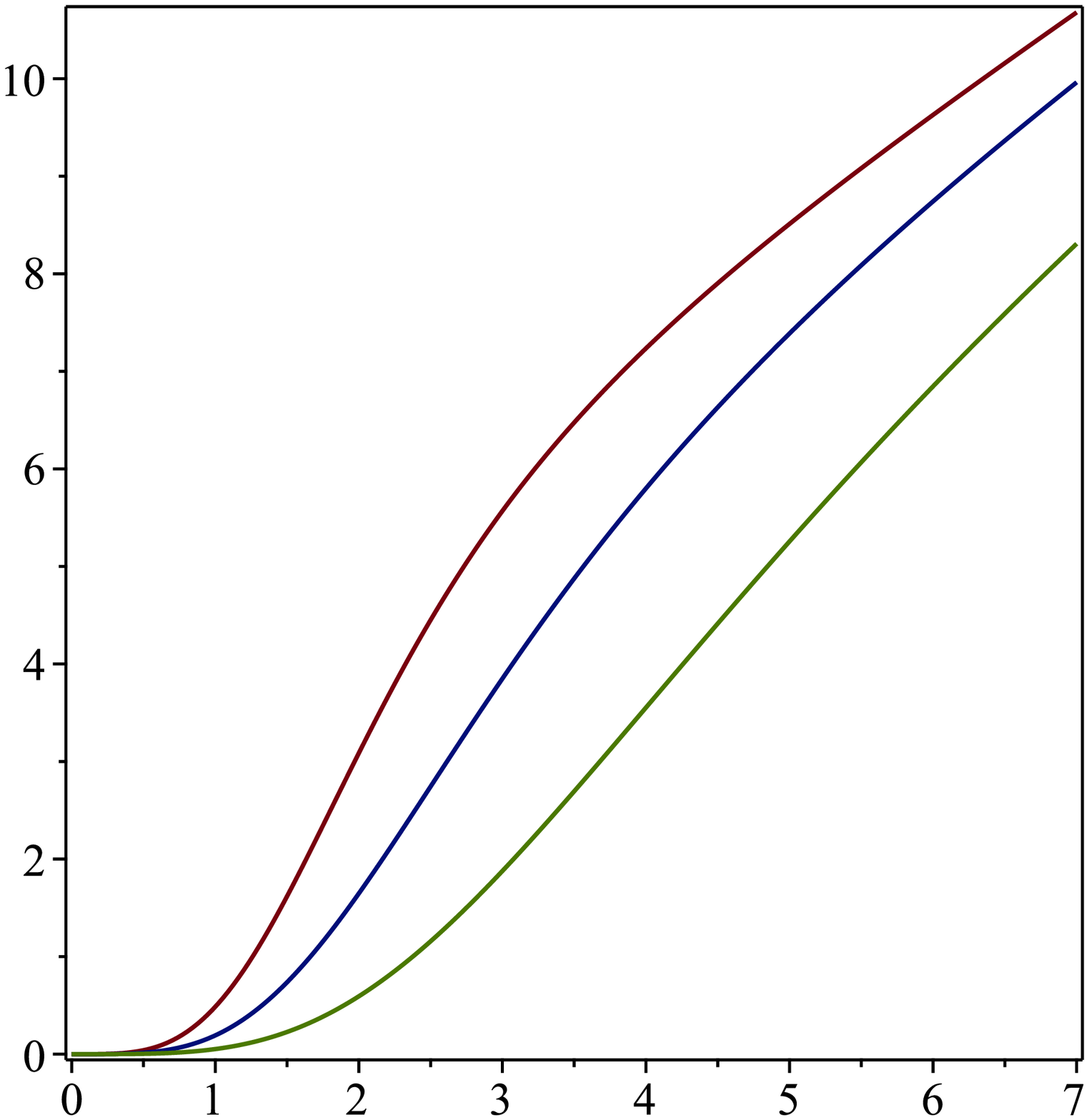}  \\  
\tilde R & \tilde R \\
  &  \\
(a)    &  (b) 
\end{array}
$$	
\caption{ $(a)$  Relativistic and Newtonian  rotation curves  $v_c^2$  (solid curves) and  $v_N^2$ (dashed curves)  for the first model of Miyamoto-Nagai-like  thick disks with parameters    $\tilde a = 1$,  $\tilde b = 0.5$ (top curves),    $\tilde b = 1$   and   $\tilde b = 2$ (bottom curves),  as function of $\tilde R$. $(b)$ The specific angular momentun $\tilde h^2$  with parameters $\tilde a =1$,   $\tilde b = 0.5$ (top curve),   $\tilde b = 1$   and $\tilde b = 2$ (bottom curve),  also as function of $\tilde R$. } 
\label{fig:fig3}
\end{figure}

 In figure \ref{fig:fig3} $(a)$    we show, as function of $\tilde R$, 
 the relativistic and Newtonian  rotation curves  $v_c^2$   and  $v_N^2$   for the first model of Miyamoto-Nagai-like  thick disks with parameters    $\tilde a = 1$,  $\tilde b = 0.5$ ,    $\tilde b = 1$   and   $\tilde b = 2$.  We  see  that  the relativistic effects  initially decrease the circular speed of the particles   and just before its maximum value is reached,  they increase it.  We also observer that such effects  become more  important in the region close to  its maximum value and when
the tangential speed is higher, according to expectations.  
Moreover, one  find that  particles become more relativistic as the distribution of matter is  flatted.  
The speed of the particles always is less than light speed  (dominant energy condition).

In figure \ref{fig:fig3}  $(b)$   we also  graph  the specific angular momentun $\tilde h^2$  for Miyamoto-Nagai-like thick disks and  the same values of parameters
 $\tilde a$  and  $\tilde b$,  also as function of $\tilde R$. We find that for these values of parameters the motion of particles is stable against radial perturbation.
 
 \section{Second family of Miyamoto-Nagai-like  thick disks}

The second Miyamoto-Nagai potential-density pair is
\begin{subequations}\begin{eqnarray}
\Phi &= &-\frac{GM } {\sqrt{R^2 + \left(a +\zeta \right)^2}} \left [  1 + \frac { a(a + \zeta ) }{ R^2 + \left(a +\zeta \right)^2 }    \right ]     ,   \label{eq:poten2}  \\
\rho_{N} &=& \frac{3 b^{2} M  \left [  R^2 (a^3 + \zeta ^3) + (a + \zeta )^3 (\zeta ^2 + 4 a \zeta + a^2)   \right ] }{  4 \pi  \zeta ^3 [ R^2 + \left(a +\zeta \right)^2]^{7/2} }, 
\end{eqnarray} \end{subequations} 
where  $\zeta = \sqrt{z^2 + b^2}$ and  again $a$ and $b$ are non-negative parameters. 
The Newtonian circular speed on $z = 0$ is given by
\begin{equation}
v_{N}^{2} = \frac{GM R^{2}\left[ R^{2}+\left( a+b\right) \left( 4a+b\right) \right] }{\left[ R^{2}+\left( a+b\right)^{2} \right]^{5/2}}.
\end{equation}

The relativistic expressions for the energy density and average pressure are
\begin{subequations}\begin{eqnarray}
\rho &= &  \frac{24 b^2 \left[ \tilde R^2+( \tilde a +\tilde \zeta)^{2} \right]^4  \left [ \tilde R^2 (\tilde a^3 + \tilde \zeta ^3) + (\tilde a + \tilde \zeta )^3 (\tilde \zeta ^2 + 4 \tilde a \tilde\zeta + \tilde a^2)   \right ]    }
{G^3 M^2 \pi \tilde \zeta ^3  \left [  2 \left (  \tilde R^2+( \tilde a +\tilde \zeta )^{2} \right )  ^{3/2} 
+ \tilde R^2+( \tilde a +\tilde \zeta )^{2}  + 
\tilde a(\tilde a +\tilde \zeta ) \right ]^5 }, \\
p &= & \frac{8\left( \tilde{R}^{2}+\left( \tilde{a}+\tilde \zeta\right)^{2} \right)^{4}\left( \tilde{z}^{2}\left[ \tilde \zeta\tilde{R}^{2}+\left( \tilde{a}+\tilde \zeta\right)^{2}\left( 3\tilde{a}+\tilde \zeta\right)  \right]^{2}+\tilde \zeta^{2}\tilde{R}^{2}\left[ \tilde{R}^{2}+\left( \tilde{a}+\tilde \zeta\right)\left( 4\tilde{a}+\tilde \zeta\right)\right]^{2}  \right)}{3\pi G^{3}M^{2} \tilde \zeta^{2} \left(\tilde{R}^{2}+\left( \tilde{a}+\tilde \zeta\right)\left( 2\tilde{a}+\tilde \zeta\right)+2\left[ \tilde{R}^{2}+\left( \tilde{a}+\tilde \zeta\right)^{2} \right]^{3/2}\right)^{6} }.  \nonumber \\
& & 
\end{eqnarray} \end{subequations} 
By inspection, we also see  that the energy density is always  
a positive quantity  in agreement with  the  weak energy condition and  
the stresses are positive  for all  values of parameters, so that we have always pressure.
On the equatorial plane $z = 0$,  the circular speed and the specific angular momentum are given by 
\begin{subequations}\begin{eqnarray}
v_{c}^{2} &= &  \frac{ 2 \tilde R^2 \left [ \tilde R^2+ (\tilde a+\tilde b)(4\tilde a+\tilde b) \right] }
{ 2 \left [  \tilde R^2+ (\tilde a+\tilde b)^2  \right]^{5/2}   + 
\left [ \tilde R^2+ (\tilde a+\tilde b)(2\tilde a+\tilde b) \right] \left [  \tilde R^2+ (\tilde a+\tilde b)^2  \right]
 -2 \tilde R ^2 \left [   \tilde R^2+ (\tilde a+\tilde b)(4\tilde a+\tilde b)  \right]  }  , \\
h^{2} &= &     \frac{   (8 \pi )^{-1} \tilde R^4  
\left [   2 \left [  \tilde R^2+ (\tilde a+\tilde b)^2  \right]^{3/2}   + \tilde R^2+ (\tilde a+\tilde b)(2\tilde a+\tilde b) \right]   
 \left [ \tilde R^2+ (\tilde a+\tilde b)(4\tilde a+\tilde b) \right]
 \left [  \tilde R^2+ (\tilde a+\tilde b)^2  \right]^{-6}  }
{   2 \left [  \tilde R^2+ (\tilde a+\tilde b)^2  \right]^{5/2}   + 
\left [ \tilde R^2+ (\tilde a+\tilde b)(2\tilde a+\tilde b) \right] \left [  \tilde R^2+ (\tilde a+\tilde b)^2 \right]
 -4 \tilde R ^2 \left [   \tilde R^2+ (\tilde a+\tilde b)(4\tilde a+\tilde b)  \right]    } .   \\
&&
\end{eqnarray} \end{subequations} 

\begin{figure}
\begin{tabular}{cc}
\includegraphics[width=0.33\textwidth]{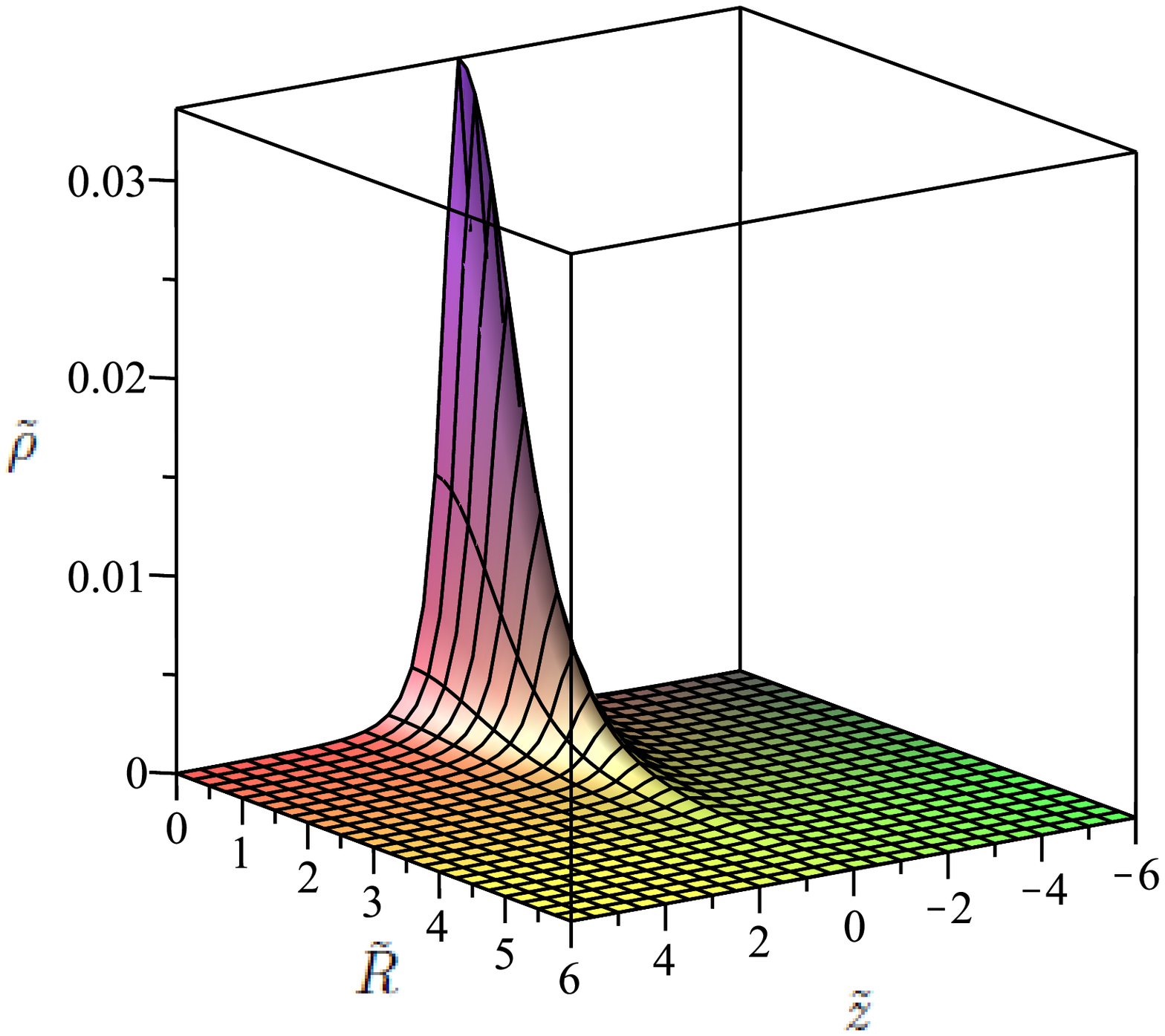} &
\includegraphics[width=0.25\textwidth]{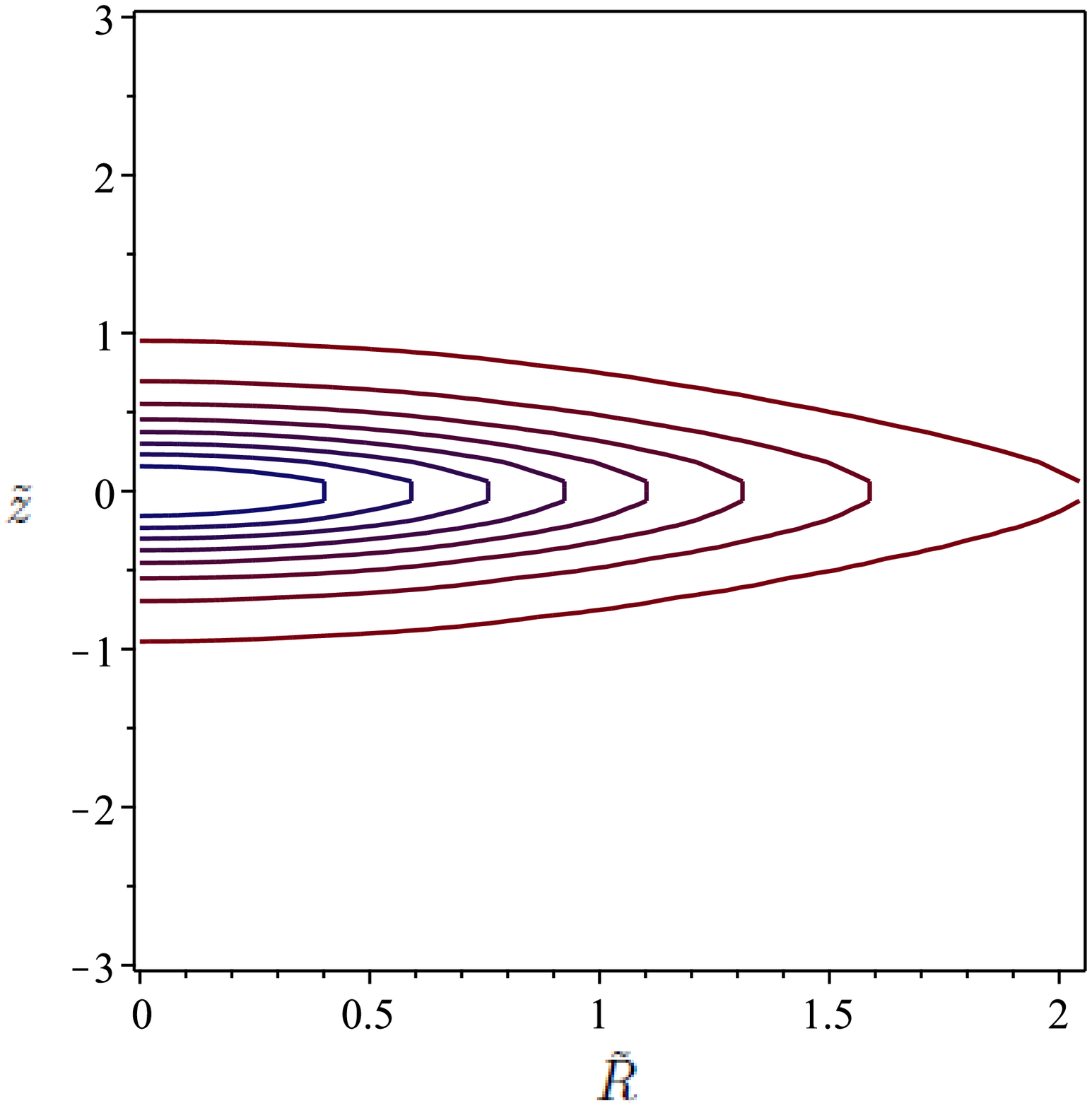}\\
&  \\
(a)    &  (b)  \\
& \\
\includegraphics[width=0.33\textwidth]{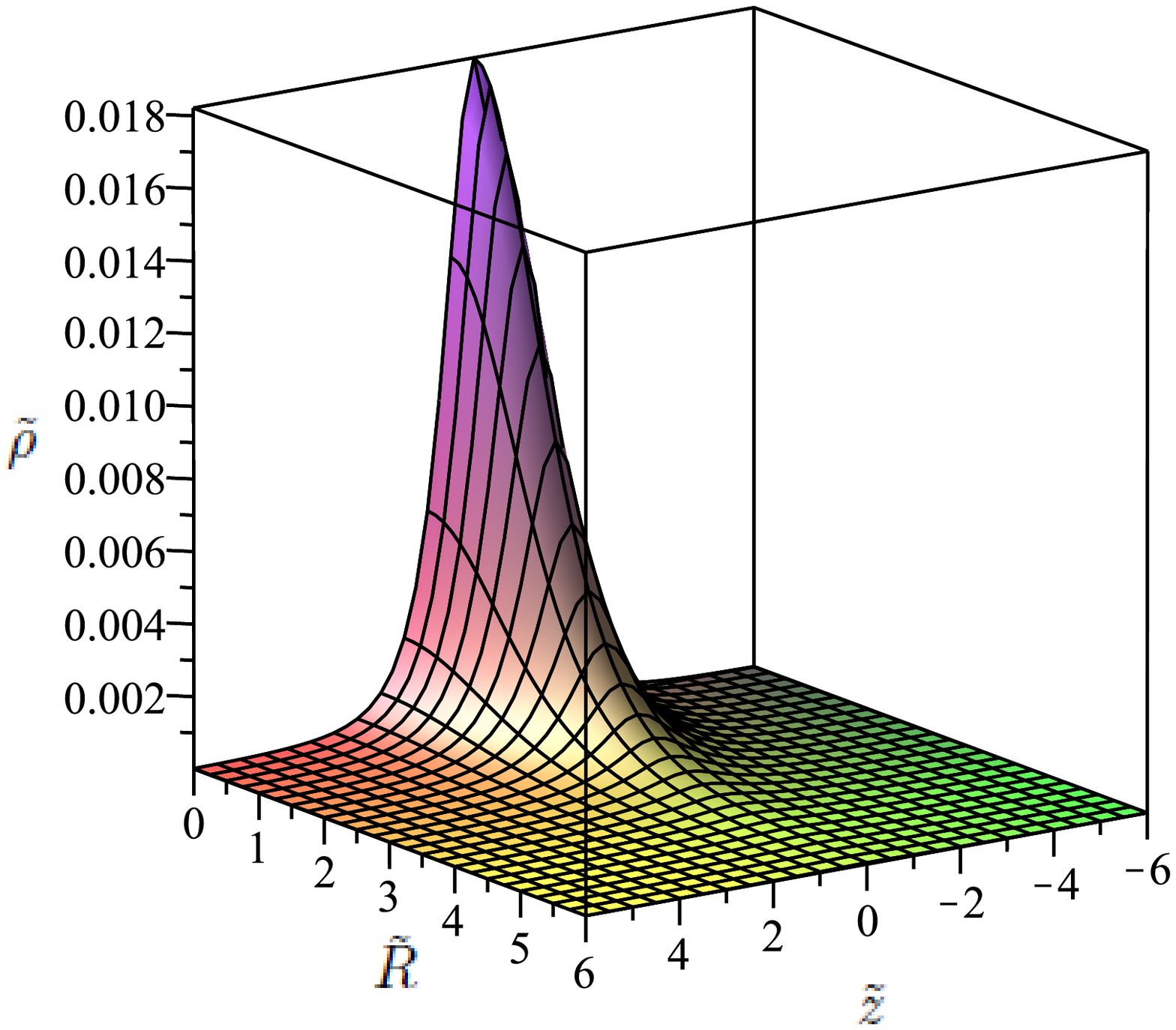} &
\includegraphics[width=0.25\textwidth]{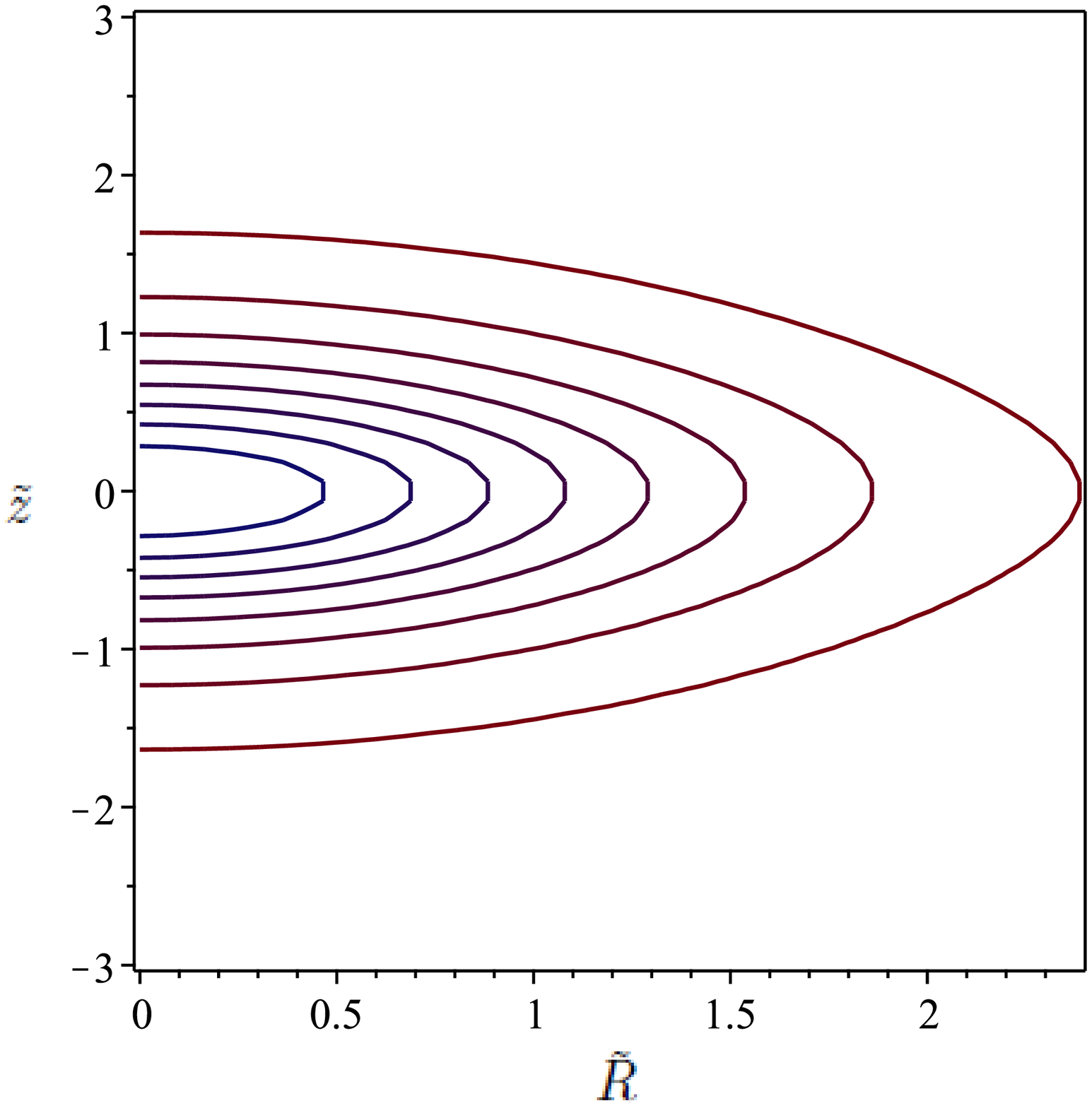}
\\
&  \\
(c)    &  (d)  \\
& \\
\includegraphics[width=0.33\textwidth]{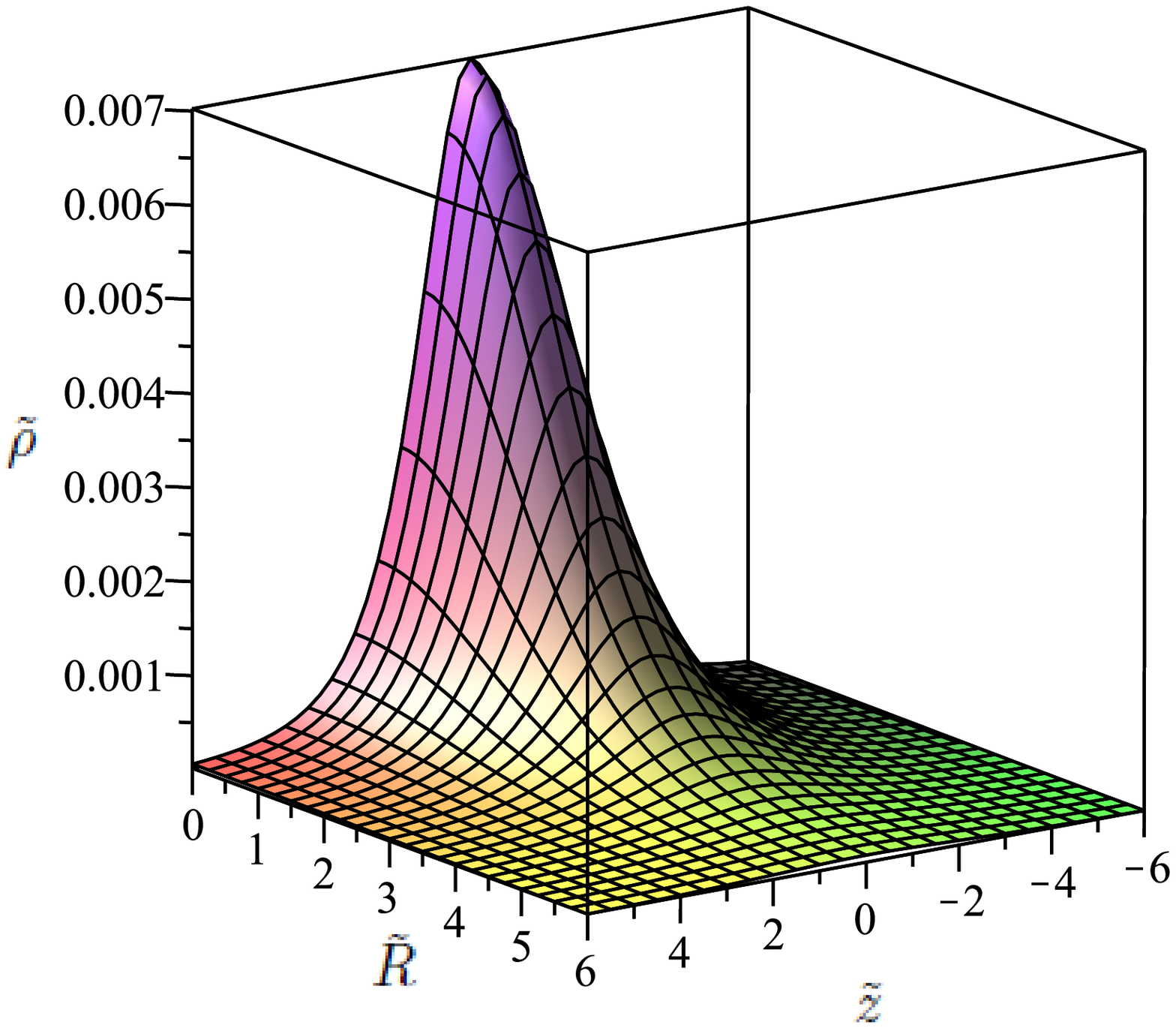} &
\includegraphics[width=0.25\textwidth]{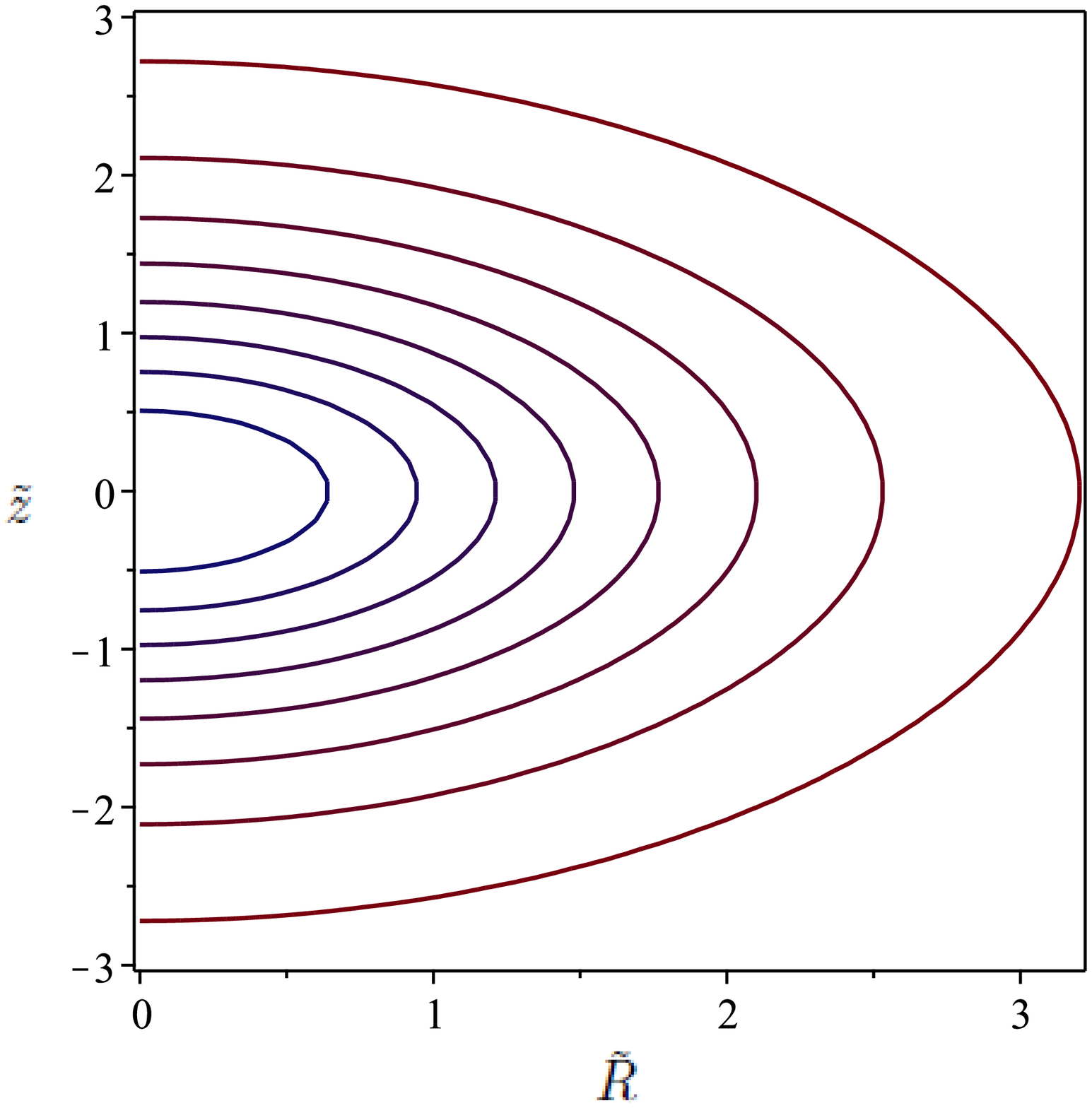}
\\
&  \\
(e)    &  (f)  \\
\end{tabular}
\caption{ The relativistic energy density $\tilde{\rho}$ and the isodensity curves for the second family  of Miyamoto-Nagai thick disk with parameters $\tilde{a} = 1$ and $\tilde{b} = 0.5$ (top figures), $1$ (middle figures), $2$ (bottom figures).}
\label{fig:fig4}
\end{figure}

\begin{figure}
\begin{tabular}{cc}
\includegraphics[width=0.33\textwidth]{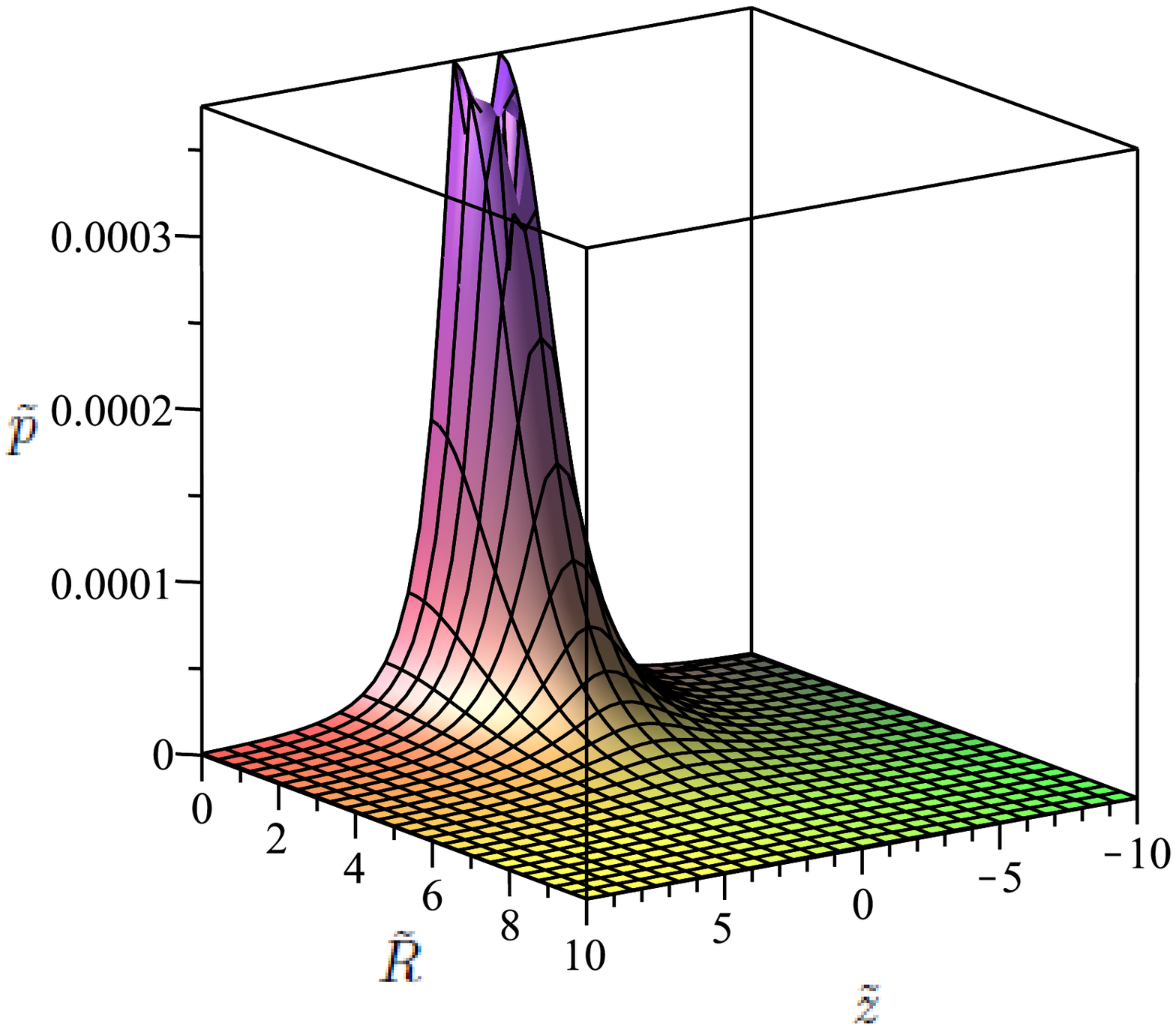} &
\includegraphics[width=0.25\textwidth]{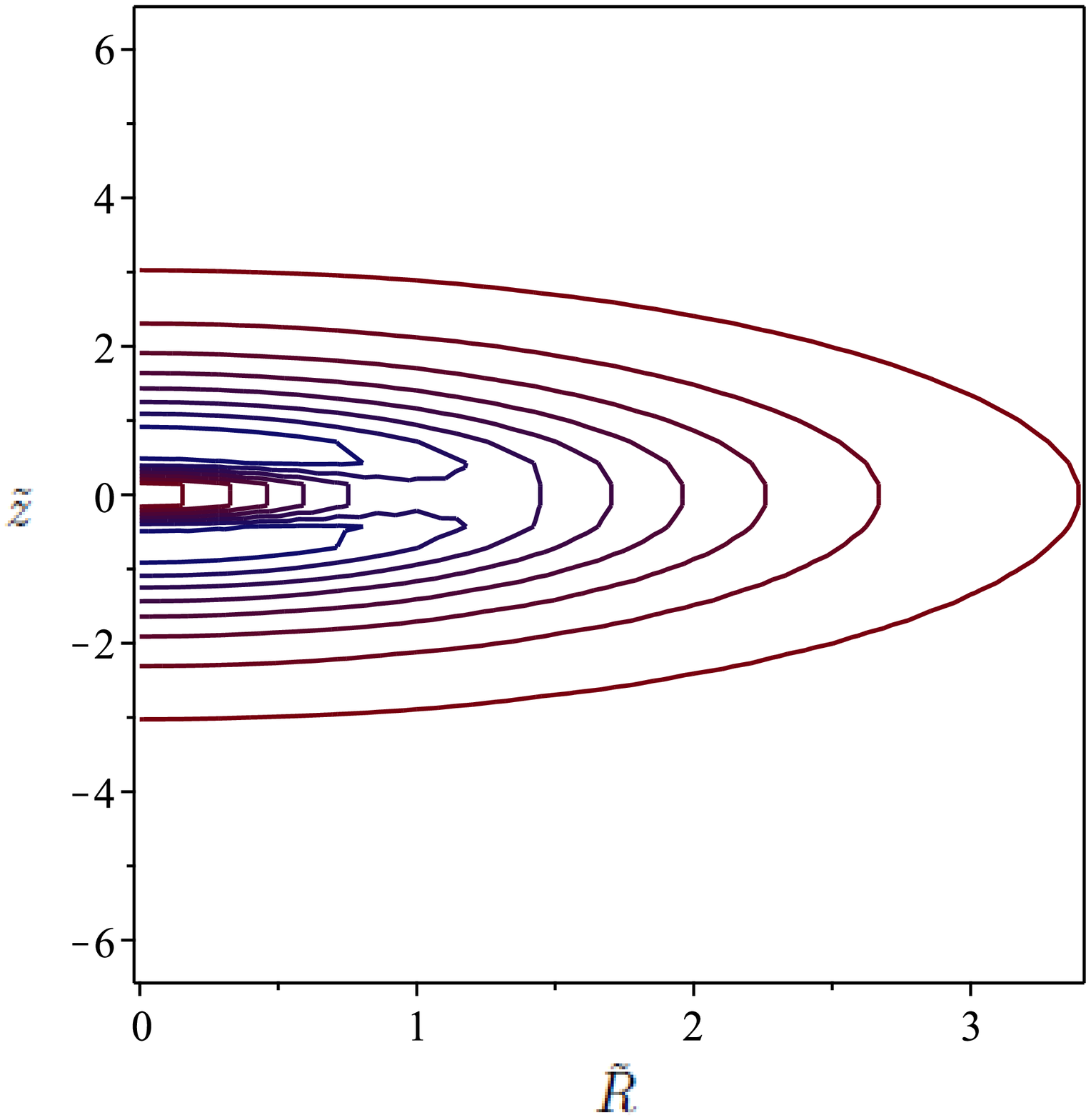}\\
&  \\
(a)    &  (b)  \\
& \\
\includegraphics[width=0.33\textwidth]{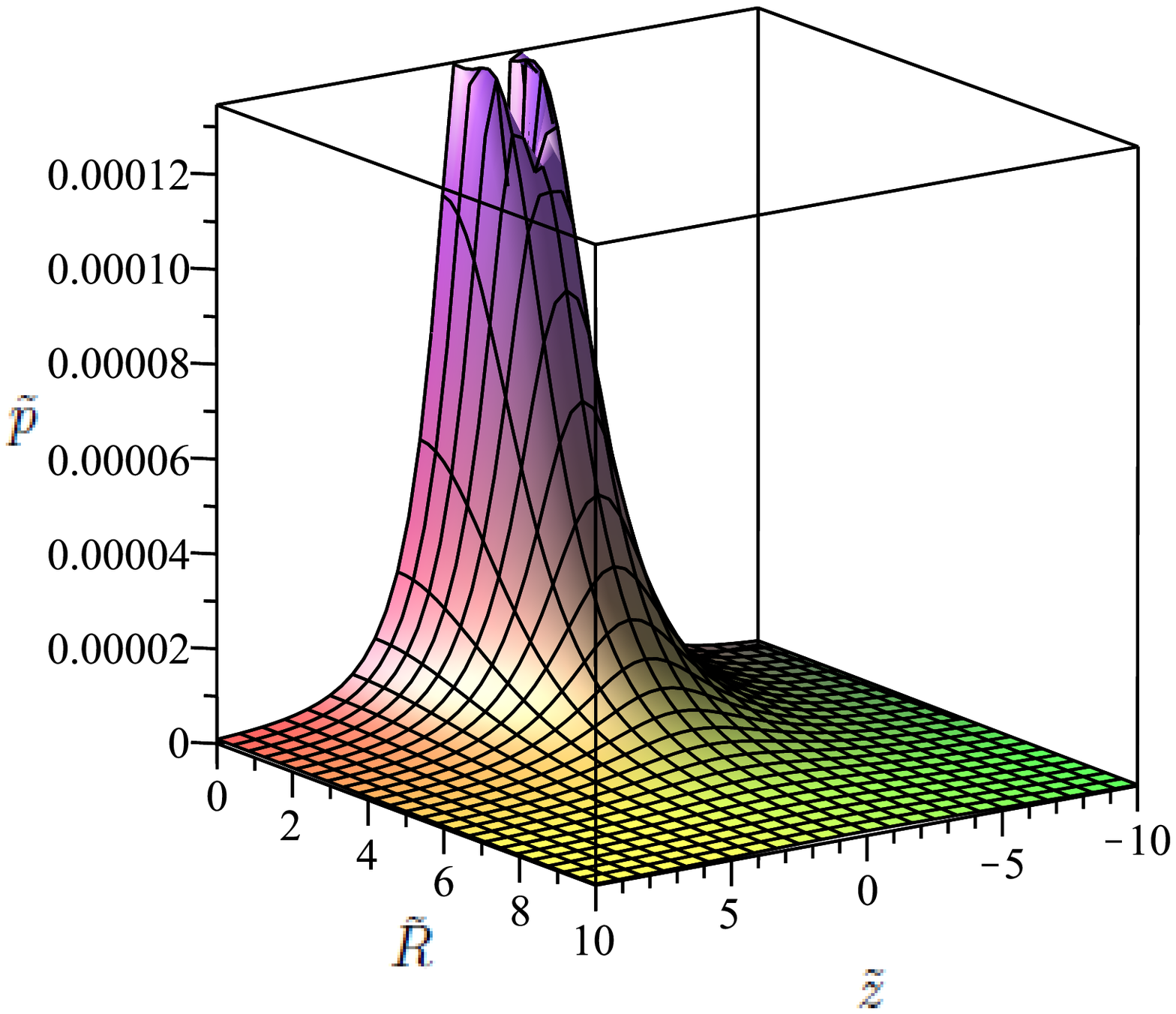} &
\includegraphics[width=0.25\textwidth]{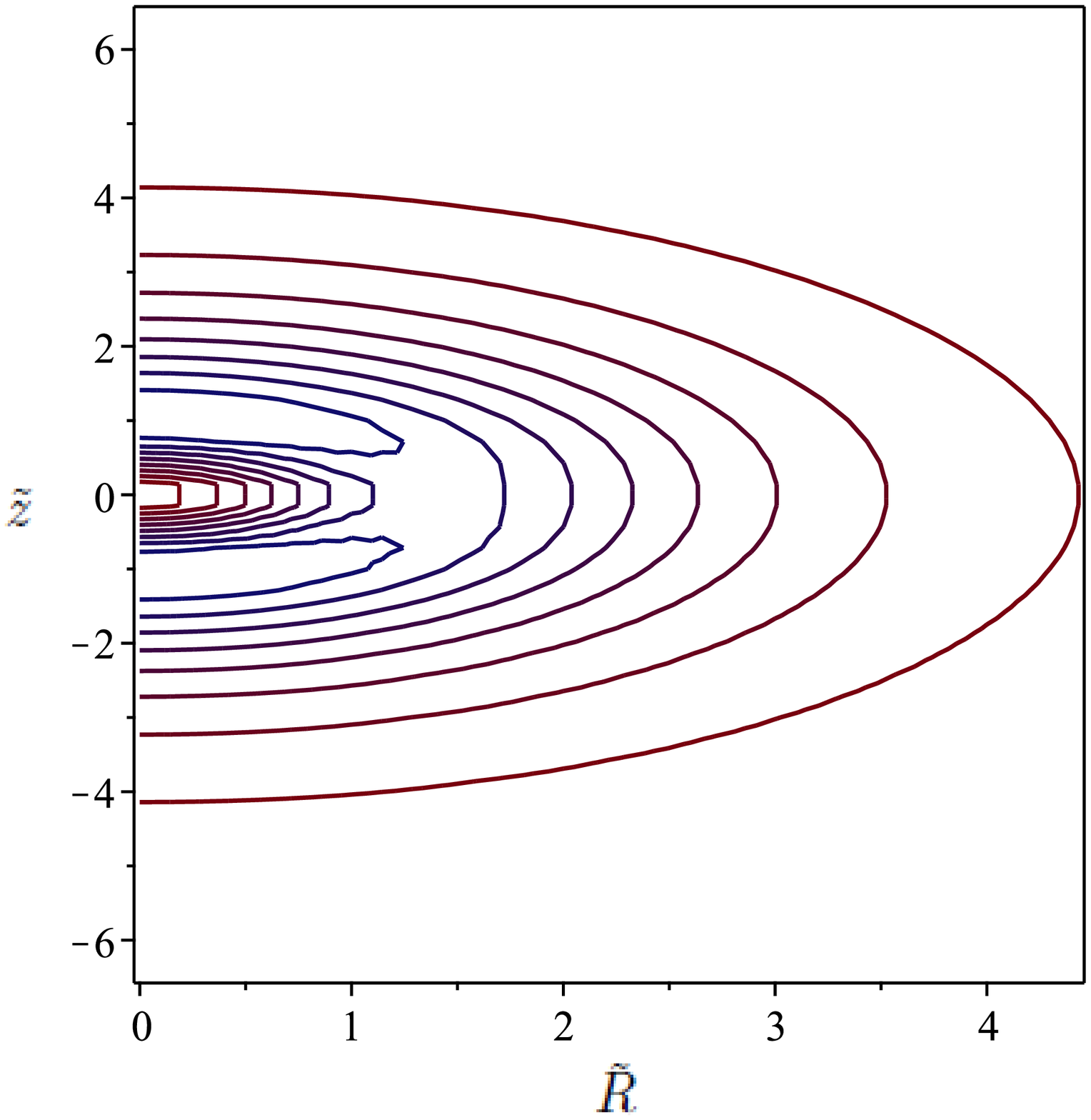}
\\
&  \\
(c)    &  (d)  \\
& \\
\includegraphics[width=0.33\textwidth]{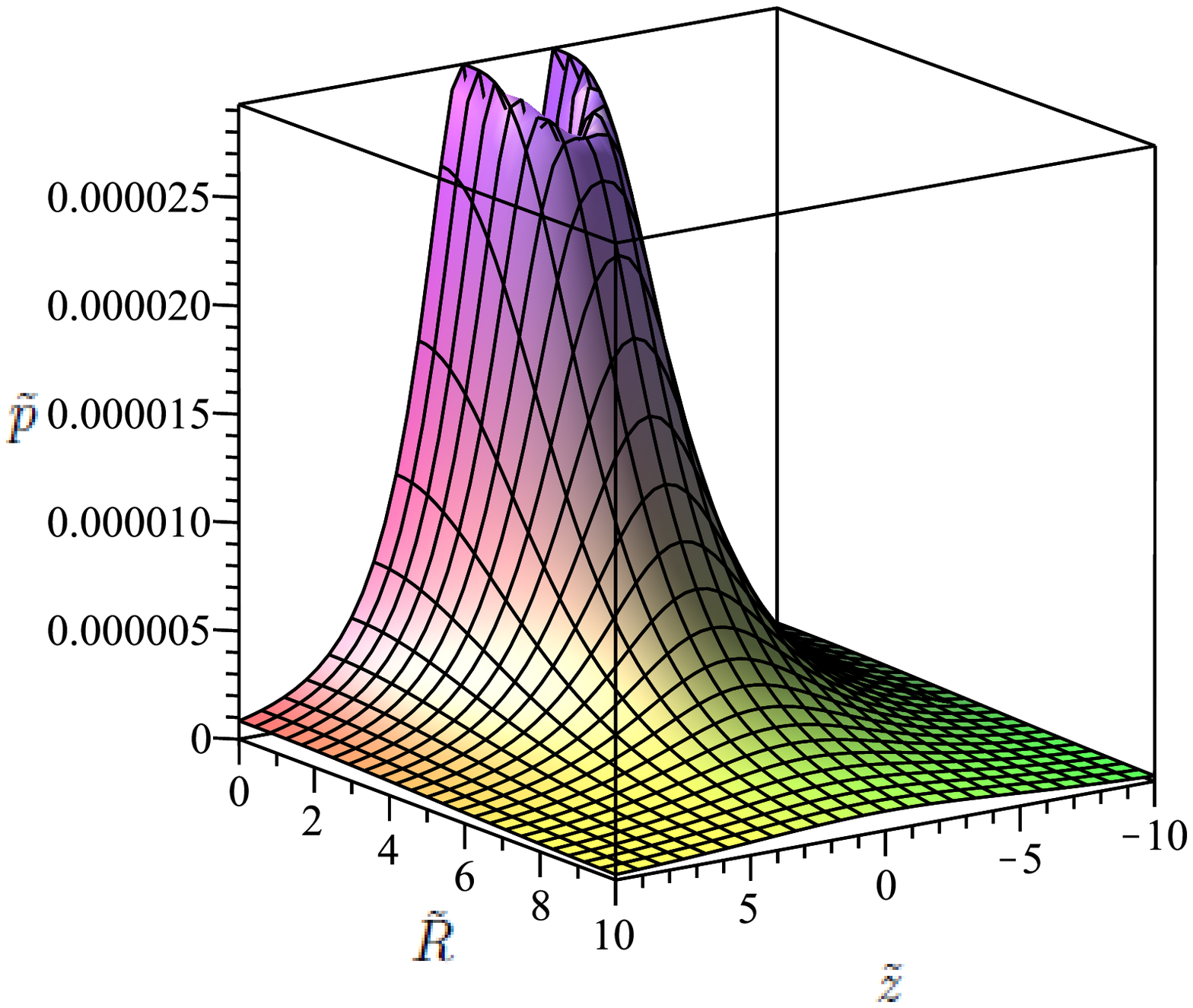} &
\includegraphics[width=0.25\textwidth]{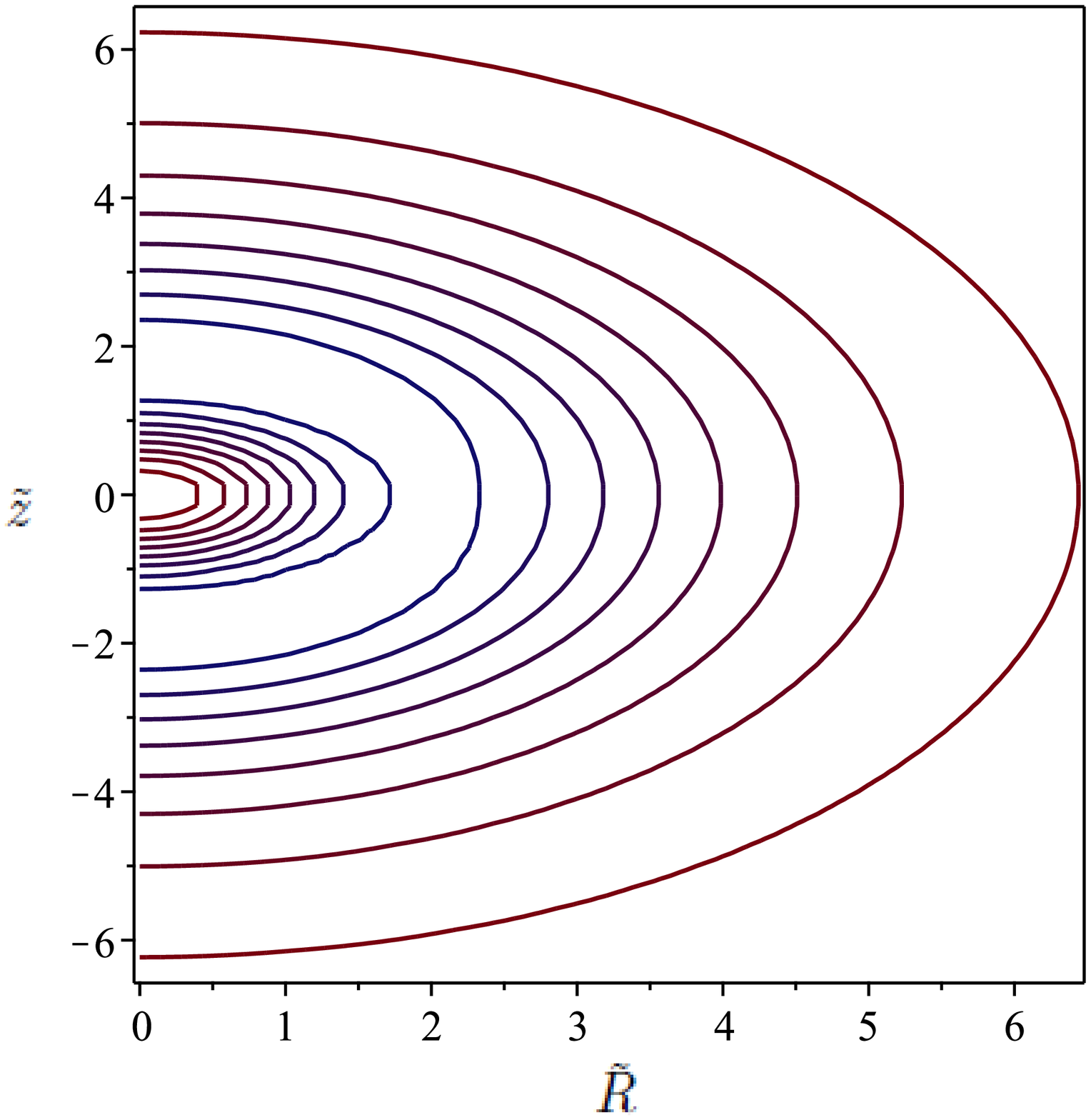}
\\
&  \\
(e)    &  (f)  \\
\end{tabular}
\caption{The relativistic average pressure $\tilde{p}$ and level curves for the second family  of Miyamoto-Nagai thick disk with parameters $\tilde{a} = 1$ and $\tilde{b} = 0.5$ (top figures), $1$ (middle figures), $2$ (bottom figures).}
\label{fig:fig5}
\end{figure}

In figures \ref{fig:fig4} and  \ref{fig:fig5} we plot, as functions of $\tilde{R}$ and $\tilde{z}$, the relativistic energy density $\tilde{\rho} = G^3M^2\rho$, the average pressure $\tilde{p} = G^3M^2p$ and, in both cases,   the level curves for the second model of Miyamoto-Nagai-like thick disks with parameters  $\tilde a = 1$,  $\tilde b = 0.5$ ,    $\tilde b = 1$   and   $\tilde b = 2$. We observer  that these functions present the same behavior as the previous model.   However, since the gravitational potential 
 (\ref{eq:poten2}) (and hence the gravitational field)
 in absolute value  is larger 
 than  the first model, these quantities have an order of magnitude higher in the second model.   

\begin{figure}
\begin{tabular}{cc}
\includegraphics[width=0.3\textwidth]{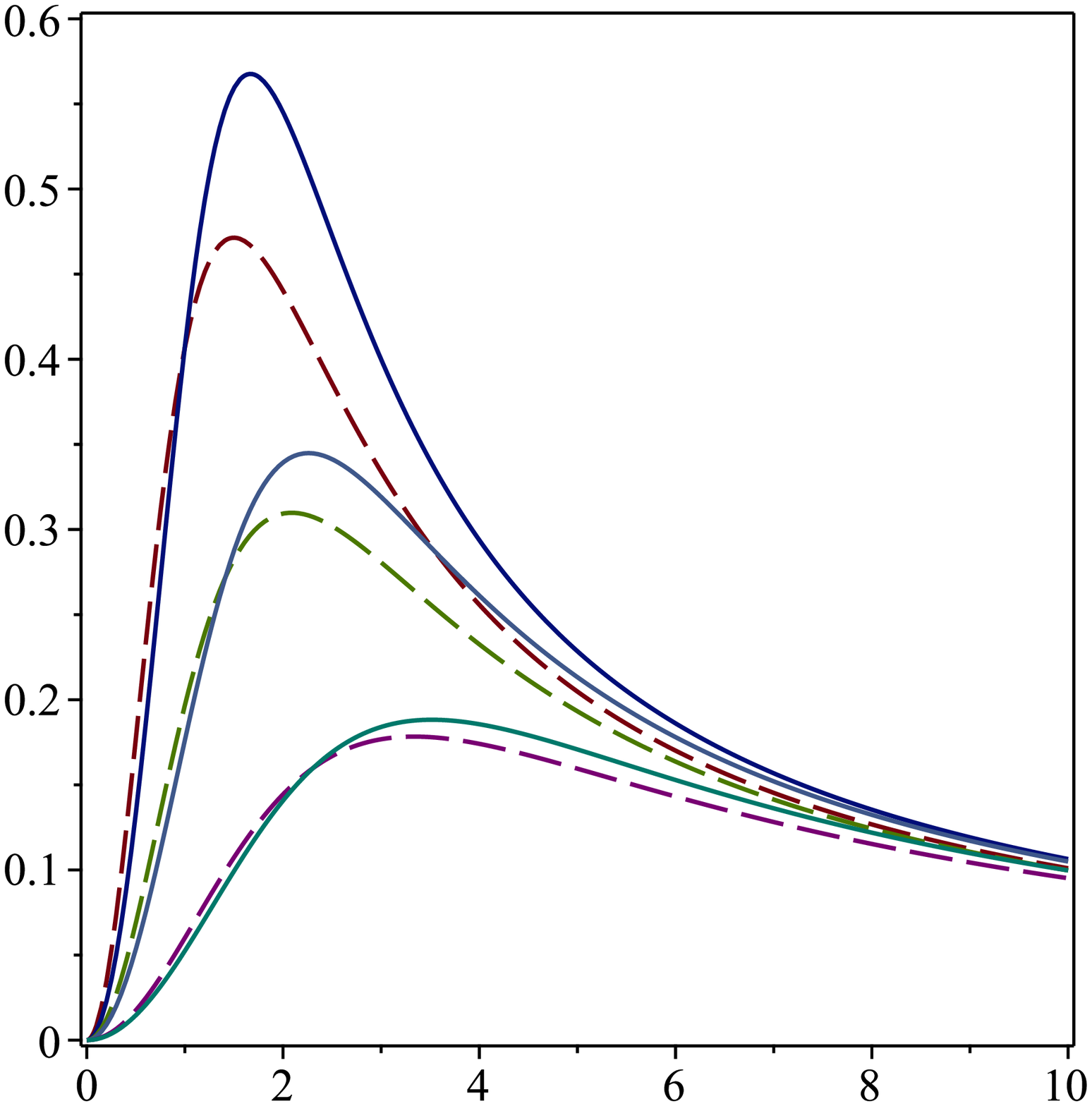} &
\includegraphics[width=0.3\textwidth]{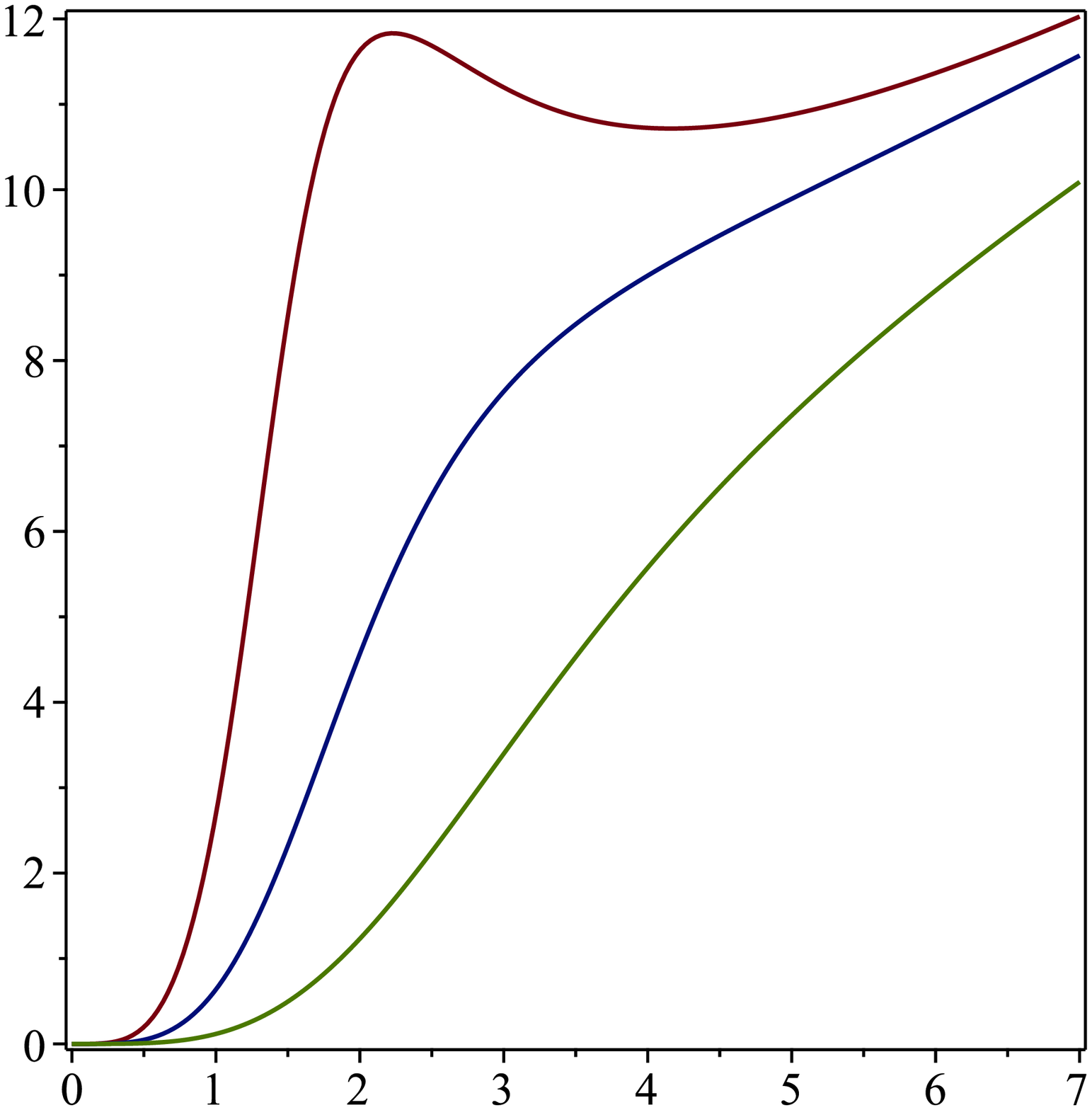}\\
&  \\
(a)    &  (b)  \\
\end{tabular}
\caption{$(a)$  Relativistic and Newtonian  rotation curves  $v_c^2$  (solid curves) and  $v_N^2$ (dashed curves)  for the second model of Miyamoto-Nagai-like  thick disks with parameters    $\tilde a = 1$,  $\tilde b = 0.5$ (top curves),    $\tilde b = 1$   and   $\tilde b = 2$ (bottom curves),  as function of $\tilde R$. $(b)$ The specific angular momentum $\tilde h^2$  with parameters $\tilde a =1$,   $\tilde b = 0.5$ (top curve),   $\tilde b = 1$   and $\tilde b = 2$ (bottom curve),  also as function of $\tilde R$. }
\label{fig:fig6}
\end{figure}

In figure \ref{fig:fig6} $(a)$  we depict, as function of $\tilde{R}$, 
the relativistic and Newtonian  rotation curves  $v_c^2$   and  $v_N^2$   for the second family  of Miyamoto-Nagai-like thick disks 
for the same value of the parameters. These quantities also show the same behavior as the first model. However,  one find    that
the relativistic effects are more significant in the second model
and moreover they  have a higher value   because  the presence of a  stronger gravitational field.
Also here,   the speed of the particles  always is less than light speed (dominant energy condition). 
 In figure \ref{fig:fig6} $(b)$  we  show the specific angular momentum $\tilde{h}^{2}$ for  the same values of parameters, also as function or $\tilde{R}$. Unlike the first model, the orbits of the particles can present regions of instability against radial perturbation   when the structures are flatted (curve $\tilde b=0.5$).

\section{A  relativistic model of galaxy }

In Newtonian gravity, the Galactic potential is usually modeled by the sum of  three  components:   a central spherical bulge  $\Phi_B$, a  thick disk  $\Phi_D$ and  a spherical dark halo $\Phi_H$, that is 
\begin{equation}
\Phi = \Phi_B + \Phi_D + \Phi_H.
\end{equation}

The bulge and disk potentials
 are  represented in the form proposed by Miyamoto
and Nagai
\begin{subequations}\begin{eqnarray}
\Phi_B &=& -\frac{GM_B}{\sqrt{r^2 + b_B^2}} ,  \\
&  & \nonumber   \\
\Phi_D &=& -\frac{GM_D}{\sqrt{R^2 + \left(a_D +\sqrt{ z^2 + b_D^2}\right)^2}} ,  \\
&  & \nonumber   
\end{eqnarray} \end{subequations}
where $r = \sqrt {R^2 + z^2}$,  and to describe the matter dark halo the  potential is assumed to be  the 
NFW model 
\begin{equation}
\Phi_H =  -  \frac{GM_H}{r}  \ln \left( 1 + \frac{r}{a_H}  \right),  
\end{equation}
where $M_H$ is the  dark halo mass and $a_H$ a scale radius. 
This potential satisfies our requirement that at large $R$ the  metric function   $\phi \ll 1$.

Since the Poisson's equation is linear, the total mass density is the sum of the components  
\begin{equation}
\rho_N = \rho_{N(B)} + \rho_{N(D)} + \rho_{N(H)}.
\end{equation}
For the above potentials we have 
\begin{subequations}\begin{eqnarray}
 \rho_{N(B)} &=&   \frac{ 3b_B^2 M_B}{4\pi  ( r^2 + b_B^2 )^{5/2} }   ,  \\
&  & \nonumber   \\
 \rho_{N(D)} &=&  \frac{ b_D^2 M_D
\left[   a_D R^2 + ( a_D + 3 \sqrt{z^2 + b_D^2 }  ) ( a_D +  \sqrt{z^2 + b_D^2 } )^2 \right]   }
{ 4\pi \left[ R^2+\left( a_D+\sqrt{z^2 + b_D^2 }\right) ^2 \right]^{5/2}    ( z^2 + b_D^2 )^{3/2} }, \\
&  & \nonumber   \\
\rho_{N(H)}   &=&     \frac{M_H}{4 \pi} \frac{1}{r (r+a_H)^2}              .
\end{eqnarray} \end{subequations}

From (\ref{eq:vn2}) it follows that  the total circular speed is also  the sum of the different contributions 
\begin{equation}
v^2_N = v^2_{N(B)} + v^2_{N(D)} + v^2_{N(H)}.
\end{equation}
For the above potentials we have 
\begin{subequations}\begin{eqnarray}
v_{N(B)}^2 &=& \frac{GM_B R^2}{\left(  R^2 +   b _B^2  \right)^{3/2} } ,  \\
&  & \nonumber   \\
 v_{N(D)}^2 &=& \frac{GM_D R^2}{\left[  R^2 +  (a_D + b_D )^2  \right]^{3/2} }, \\
&  & \nonumber   \\
v_{N(H)}^2 &=&   v_0^2  \left [  \frac { \ln ( 1 + R/a_H)  } {R/a_H} - \frac{1}{  1 + R/a_H}             \right ] , 
\end{eqnarray} \end{subequations}
where $ v_0^2 = GM_H/a_H $. 

Thus, the total relativistic circular speed is given by the non-linear expression
\begin{equation}
v_c^2 = \frac {  v^2_{N(B)} + v^2_{N(D)} + v^2_{N(H)}}  {1-  \frac{1}{2}( \Phi_B + \Phi_D + \Phi_H  ) -   v^2_{N(B)} - v^2_{N(D)} - v^2_{N(H)}   }.
\end{equation}

\begin{figure}
\begin{tabular}{c}
\includegraphics[width=0.5\textwidth]{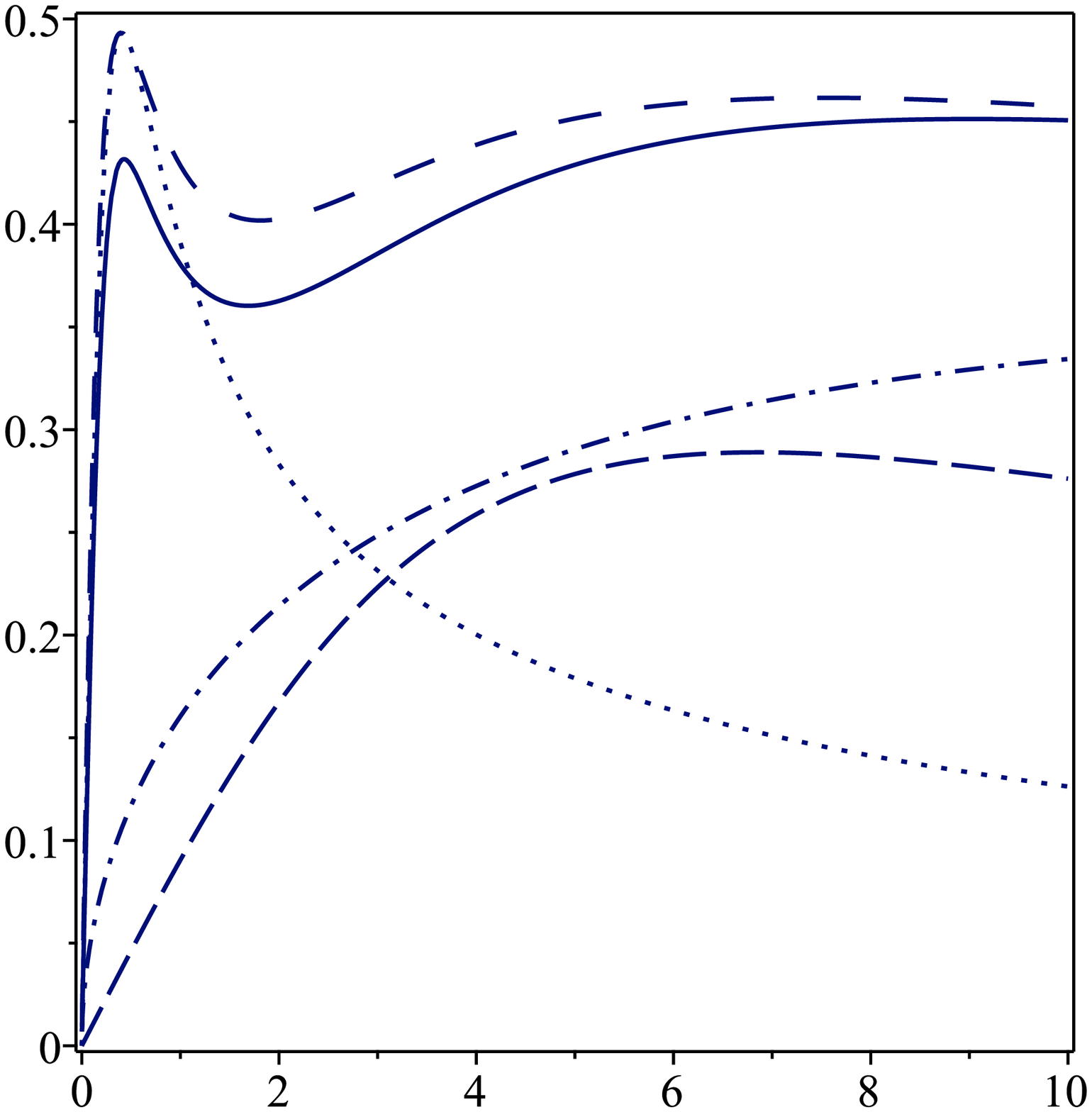} \\
 \\
 $\tilde R$    
\end{tabular}
\caption{Relativistic and Newtonian total  rotation curves 
$v_c$ (solid curve) and  $v_N$ (top curve)  and the
relativistic contributions of  the three components: central bulge $v_B$ (dotted curve), disk $v_D$ (dashed curve) and dark halo $v_H$ (dash-dotted curve), as function of $\tilde R$.}
\label{fig:fig7}
\end{figure}

In order to  analyze   total relativistic tangential velocity $v_c$   and the different components we perform a graphical analysis of them  for the values of the  potential parameters computed  in reference \cite{Bajkova} $a_D=4.4$,   $a_H=7.7$, 
  $b_B=0.2672$,  $b_D=0.3084$, $M_b=443$,  $M_d=2798$
 and  $M_h=12474 $. In figure \ref{fig:fig7}  we graph, as function of $\tilde R = R / G M_D$,  the relativistic and Newtonian total rotation curves $v_c$ and $v_N$  and
 the relativistic contributions of the three components: the central bulge  $v_B$, the disk   $v_D$ and the dark matter halo
  $v_H$. We see that   total rotation curves are  flatted after certain value of $\tilde R$
  as the observational data reveal, whereas  tangential speed of the visible matter (central bulge plus thick disk )
rapidly falls to zero.  
 We find
that the relativistic effects decrease the total  tangential velocity everywhere  and  are more significant in the region close to its maximum  value and, as expected,  for  velocities comparable to the speed of light (relativistic particles).  In fact,  
the tangential velocity  of the stars around  a typical galaxy   is about   200 - 300 km/s and therefore the relativistic effects are expected to be small.
In the case of our Galaxy for a radius of   $R=0.4$ kpc  which corresponds to a  tangential velocity about $260.8$   km/s we find that   the relativistic corrections  
 are of the order of    $1.2  \times 10^{-4}$ km/s.  However, we believe that 
 for a sufficiently large distance traveled by a star they  could become important. 
 
 \section{Discussion}

Three-dimensional axisymmetric relativistic sources  for a Majumdar-Papapetrou type conformastatic
spacetime were construct    from  Newtonian potential-density pairs. As  simple examples,  were considered   two family of    Miyamoto-Nagai-like thick disk models  and a  relativistic  model
of galaxy  composite  by three components: a central bulge, a disk and a NFW dark matter halo.  
   
The equatorial
circular motion of test particles around such configurations were studied,  and in all models  was observed  that  the relativistic effect are more important in the regions around  its maximum value  and, as expected, for velocities comparable to the speed of light. 
In the case of the thick disks was found that   such effects  initially decrease the circular speed of the particles   and just before reaching  its maximum value,  they increase it, whereas for the relativistic model  of galaxy presented 
 the relativistic effects always  decrease the circular speed.  
Also was  observed    that
the relativistic corrections are more significant in the second  disk model. Moreover, all the physical quantities computed have an order of magnitude higher in the second model due to the presence of a  stronger gravitational field. 

The stability  of the
orbits  for radial perturbation  was analyzed  for the thick disks using an extension of the Rayleigh criterion. 
We found  values of parameters for which  the motion of particles is stable against radial perturbation.
The models constructed  satisfying all the energy conditions.



\section*{References}

\begin{thebibliography}{999} 


\bibitem{KUZMIN} G. G. Kuzmin 1956,  Astron. Zh.  {\bf 33}, 27 (1956). 

\bibitem{TOOMRE} A. Toomre, Ap. J.  {\bf 138}, 385  (1962). 

\bibitem{Miyamoto}  M.  Miyamoto and   R. Nagai, PASJ  {\bf 27}, 533 (1975).

\bibitem{Nagai} R. Nagai and M.  Miyamoto,  PASJ {\bf 28}, 1 (1976).

\bibitem{Plummer} H. C. Plummer,  MNRAS {\bf 71}, 460 (1911).

\bibitem{Santillan} C. Allen and A.  Santillan,   Rev. Mexicana  Astron.  Astrof. {\bf 22}, 255 (1991).

\bibitem{Nink1} S. Ninkovic, Rev. Mexicana  Astron.  Astrof. {\bf 53},  113 (2017).

\bibitem{Bajkova} A. T. Bajkova and V. V. Bobylev,   Astronomy Letters  {\bf 42},  No. 9, 567 (2016).

\bibitem{Nink2}  S. Ninkovic, PASA  {\bf 32}, e032, 	(2015). 

\bibitem{Binney} J. Binney  and  S. Tremaine S, 2008, Galactic Dynamics, 2nd edn. Princeton
Univ. Press, Princeton, NJ.



\bibitem{BS} W. A. Bonnor and A. Sackfield, Commun. Math.   Phys. {\bf 8}, 338
(1968).

\bibitem{MM1} T. Morgan and L. Morgan, Phys. Rev.  {\bf 183},  1097 (1969).

\bibitem{MM2} L. Morgan and T. Morgan, Phys. Rev. D  {\bf 2},  2756 (1970).



\bibitem{LP} D. Lynden-Bell and S. Pineault, Mon. Not. R.  Astron. Soc. {\bf
185}, 679 (1978). \label{bib:LP}

\bibitem{CHGS} A. Chamorro, R. Gregory, and J. M. Stewart, Proc. R. Soc. London
{\bf A413}, 251 (1987).

\bibitem{LO} P.S. Letelier and S. R. Oliveira, J. Math.  Phys.  {\bf 28}, 165
(1987).

\bibitem{LEM} J. P. S. Lemos, Class. Quantum Grav. {\bf 6}, 1219 (1989).

\bibitem{BLK} J. Bi\u{c}\'{a}k, D. Lynden-Bell, and J.  Katz,  Phys. Rev. D {\bf
47}, 4334 (1993).

\bibitem{BLP} J. Bi\u{c}\'{a}k, D. Lynden-Bell, and C.  Pichon, Mon. Not. R.
Astron. Soc. {\bf 265}, 126 (1993).

\bibitem{GE} G. A. Gonz\'alez and O. A. Espitia,  Phys. Rev. D  {\bf 68}, 104028
(2003).



\bibitem{BL} J. Bi\u{c}\'ak and T. Ledvinka, Phys. Rev.  Lett. {\bf 71}, 1669
(1993).

\bibitem{GL2} G. A. Gonz\'alez and P. S. Letelier, Phys.  Rev.  D {\bf 62},
064025 (2000).


\bibitem{LL1} J. P. S. Lemos and P. S. Letelier, Class.  Quantum Grav. {\bf
10}, L75 (1993).

\bibitem{LL2} J. P. S. Lemos and P. S. Letelier, Phys. Rev. D  {\bf 49},  5135
(1994).


\bibitem{G-L-thick}  G. A. Gonz\'alez and P. S. Letelier, Phys.  Rev.  D {\bf 69}, 044013 (2004).

\bibitem{Let-galaxy}  D.  Vogt and P. S.  Letelier, MNRAS {\bf 363}, 268 ( 2005).

\bibitem{Gon-shell}  G.  Garc\'{\i}a-Reyes,  Gen. Relativ. Gravit. {\bf 49}, 3, 1 (2017).

\bibitem{NFW}  J. F.  Navarro, C. S. Frenk and S. D. M. White, Ap. J. {\bf  462}, 563  (1996). 


\bibitem{RAYL} Lord Rayleigh, 1917, Proc. R. Soc. London A, 93, 148

\bibitem{FLU} L.  D. Landau and E. M. Lifshitz, {\it Fluid 
Mechanics}(Addison-Wesley, Reading, MA, 1989).


\bibitem{synge}  J. L. Synge, Relativity: The General Theory (North-
Holland, Amsterdam, 1966).

\bibitem{kramer} H. Stephani, D. Kramer, M. McCallum, C. Hoenselaers,
and E. Herlt, Exact Solutions of Einsteins’s Field Equations
(Cambridge University Press, Cambridge, England, 2003).

\bibitem{majum} S. D. Majumdar,  Phys. Rev. {\bf 72},  390 (1947).

\bibitem{papa} A. Papapetrou , Proc. Roy. Soc. (London)  {\bf A51}, 191 (1947).

\end{thebibliography}
\end{document}